\documentclass[twocolumn]{aastex62}

\usepackage{lineno}
\graphicspath{{./}{figures/}}

\received{September 16, 2019}
\revised{November 18, 2019}
\accepted{December 16, 2019}
\published{\href{https://doi.org/10.3847/1538-4365/ab6324}{February 3, 2020 in ApJS}}
\shorttitle{Imaging of streamers with \textit{PSP}}
\shortauthors{Poirier et al.}
\begin{document}

\title{Detailed imaging of coronal rays with the \textit{Parker Solar Probe}}

\author[0000-0002-1814-4673]{Nicolas Poirier}
\email{npoirier@irap.omp.eu}
\affiliation{IRAP, Universit\'e Toulouse III - Paul Sabatier,
CNRS, CNES, Toulouse, France}

\author{Athanasios Kouloumvakos}
\affiliation{IRAP, Universit\'e Toulouse III - Paul Sabatier,
CNRS, CNES, Toulouse, France}

\author[0000-0003-4039-5767]{Alexis P. Rouillard}
\affiliation{IRAP, Universit\'e Toulouse III - Paul Sabatier,
CNRS, CNES, Toulouse, France}

\author[0000-0001-8247-7168]{Rui F. Pinto}
\affiliation{IRAP, Universit\'e Toulouse III - Paul Sabatier,
CNRS, CNES, Toulouse, France}

\author[0000-0002-8164-5948]{Angelos Vourlidas}
\affiliation{John Hopkins APL, Laurel, USA}

\author[0000-0001-8480-947X]{Guillermo Stenborg}
\affiliation{Naval Research Laboratory, Washington DC, USA}

\author{Emeline Valette}
\affiliation{IRAP, Universit\'e Toulouse III - Paul Sabatier,
CNRS, CNES, Toulouse, France}

\author[0000-0001-9027-8249]{Russell A. Howard}
\affiliation{Naval Research Laboratory, Washington DC, USA}

\author[0000-0003-1377-6353]{Phillip Hess}
\affiliation{Naval Research Laboratory, Washington DC, USA}

\author{Arnaud Thernisien}
\affiliation{Naval Research Laboratory, Washington DC, USA}

\author{Nathan Rich}
\affiliation{Naval Research Laboratory, Washington DC, USA}

\author[0000-0001-8956-2824]{L\'ea Griton}
\affiliation{IRAP, Universit\'e Toulouse III - Paul Sabatier,
CNRS, CNES, Toulouse, France}

\author{Mikel Indurain}
\affiliation{IRAP, Universit\'e Toulouse III - Paul Sabatier,
CNRS, CNES, Toulouse, France}

\author{Nour-Edine Raouafi}
\affiliation{John Hopkins APL, Laurel, USA}

\author[0000-0001-6216-6530]{Michael Lavarra}
\affiliation{IRAP, Universit\'e Toulouse III - Paul Sabatier,
CNRS, CNES, Toulouse, France}

\author[0000-0002-2916-3837]{Victor R\'eville}
\affiliation{IRAP, Universit\'e Toulouse III - Paul Sabatier,
CNRS, CNES, Toulouse, France}



\begin{abstract}
The Wide-field Imager for Solar PRobe (WISPR) obtained the first high-resolution images of coronal rays at heights below 15~\textit{R}$_\odot$ when \textit{Parker Solar Probe} (\textit{PSP}) was located inside 0.25~au during the first encounter. We exploit these remarkable images to reveal the structure of coronal rays at scales that are not easily discernible in images taken from near 1~au. To analyze and interpret WISPR observations, which evolve rapidly both radially and longitudinally, we construct a latitude versus time map using the full WISPR dataset from the first encounter. From the exploitation of this map and also from sequential WISPR images, we show the presence of multiple substructures inside streamers and pseudostreamers. WISPR unveils the fine-scale structure of the densest part of streamer rays that we identify as the solar origin of the heliospheric plasma sheet typically measured in situ in the solar wind. We exploit 3D magnetohydrodynamic models, and we construct synthetic white-light images to study the origin of the coronal structures observed by WISPR. Overall, including the effect of the spacecraft relative motion toward the individual coronal structures, we can interpret several observed features by WISPR. Moreover, we relate some coronal rays to folds in the heliospheric current sheet that are unresolved from 1~au. Other rays appear to form as a result of the inherently inhomogeneous distribution of open magnetic flux tubes. 
\end{abstract}

%, thus providing new insights on the structure of the corona
%the time and spatial dependence of the rays imaged by a spacecraft moving rapidly both radially and in heliographic longitude, we map the WISPR data in latitude versus time.
%form a powerful tool to analyse WISPR data, including the effect of the spacecraft super-rotating towards the individual structures that make up the streamer rays. 



\section{Introduction} \label{sec:intro}

Coronal rays are narrow lanes of enhanced brightness that structure the solar corona observed in visible light. They have been contemplated for millennia during the rare and brief solar eclipses and scrutinized more systematically in the last decades with the advent of orbiting white-light (WL) coronagraphs and heliospheric imagers. We know that coronal rays correspond to electron density enhancements extending far out to several tens of solar radii (\textit{R}$_\odot$) where the solar wind has already formed \citep{Druckmuller2014ApJ}. Their appearance does not vary significantly over hourly timescales \citep[e.g.][]{Fisher1995AAS}. They stand as quiescent features relative to the ``active'' corona that is continually perturbed by transients such as coronal mass ejections (CMEs) and coronal jets. Coronal rays have been therefore associated with the quiet regions where the solar wind forms. Recent research based on high-cadence imaging of these structures reveals that their apparent quiescent nature at small spatial and hourly temporal scales is likely misleading and streamer rays could actually result from an incessant reconfiguration of their coronal source regions \citep{DeForest2018}.\\

The appearance of coronal rays evolves dramatically on daily and yearly timescales. Their daily evolution is mostly due to the effect of solar rotation and small-scale evolution of the magnetic field that brings rays located at different longitudes (and latitudes) into the plane of the sky of the observing telescope. On longer timescales, the positions and brightness of coronal rays respond to changes in the topology of the coronal magnetic field during the solar cycle \citep{Golub2009}. Despite decades of observations, the physical mechanisms that produce these rays are still debated. A source of difficulty resides in the nature of the observations themselves; any WL image of the solar corona results necessarily from the integration of sunlight that has been scattered by electrons situated along each line of sight (LOS) of each pixel in the image. \\

This observational constraint complicates any interpretation of the 3D structure of streamer rays and the determination of their source closer to the surface of the Sun. An example relates to the nature of polar rays observed in coronagraphs and eclipse images to extend over the northern and southern polar coronal holes tens of solar radii in heliocentric radial distance \citep[e.g.][]{Fisher1995AAS}. They have been associated with polar coronal holes and later reinterpreted as standard streamer rays rooted at much lower latitudes above active regions \citep{Li2000ApJ}. Recent studies have also revealed the presence of high-latitude streamers above polar-crown filaments that form on the periphery of polar coronal holes during elevated solar activity \citep{Zhukov2008ApJ}. WL plumes unambiguously related to coronal holes are also clear features in WL coronagraphs; they can extend up to many solar radii above the limb of the Sun and also contribute to the occurrence of polar rays \citep[][]{Wang1994}. \\

Continuous observations of coronal rays from multiple vantage points provided by the \textit{Solar and Heliospheric Observatory} \citep[\textit{SOHO}:][]{Domingo1995} and the \textit{Solar-Terrestrial Relations Observatory} \citep[\textit{STEREO}:][]{Kaiser2008}, combined with numerical modeling of the solar corona, have provided a global picture of the 3D topology of streamer rays that represent a subset of the brightest coronal rays. These studies have confirmed a long-recognized relationship between the 3D distribution of coronal rays and the interplanetary sector boundaries \citep[e.g. see][]{Hansen1974GeoRL,Howard1974SoPh,Wang1997APJ,Wang2000GeoRL,Liewer2001JGR,Saez2005A&A,Thernisien2006ApJ}. At the Sun, the interplanetary sector boundaries map back to coronal locations where the polarity of the solar magnetic field lines flips. This occurs in the region where the magnetic field becomes radial and can be considered locally open \citep{Smith1978JGR}. These field lines are connected to the interplanetary medium and coronal plasma flows along them, creating a heliospheric plasma sheet (HPS) around the heliospheric current sheet (HCS). The brightness of the HPS depends on its LOS depth, which in turn is determined by the structure of the HCS \citep[e.g.][]{Wang2000GeoRL}. Bright coronal rays or ``streamer rays'' mark locations of folds of the HPS or locations where the HPS is parallel to the LOS. In other words, the coronal rays provide a map of the HPS structure.\\

However, not all coronal rays are associated with polarity inversion lines (PILs). Some map back to unipolar regions. These unipolar streamers are called ``pseudostreamers'' \citep[][]{Wang2007ApJ}. While both streamers and pseudostreamers contribute to the brightness of the K-corona, only helmet streamers are associated with regions where magnetic field lines of opposite polarities meet in the interplanetary medium. A mechanism for the formation of dense plasma flows along streamers and pseudostreamers would be strong plasma heating at the base of flux tubes. Due to the presence of active regions and generally intense magnetic fields in the active region belt, the bases of streamers are typically associated with hotter plasma. These regions would drive more chromospheric evaporation, expelling denser plasma into the escaping wind \citep{Wang1994,Pinto2017ApJ}. In addition, the magnetic flux tubes that form the (true) streamers tend to reconverge near the streamer tops, potentially contributing as well to the high-density wind \citep{Wang1994}. Transition of dense plasma confined in coronal loops to open flux tubes through footpoints exchange is another possible scenario \citep[e.g.][]{Culhane2014SoPh}. The study of bright coronal features and their variability therefore provides insights into the mechanism at the origin of the formation of the slow solar wind, which is still a matter of debate.\\

Because of the LOS effects, it is generally difficult to analyze the detailed streamer topology from 1~au. For this reason, the \textit{Parker Solar Probe} \citep[\textit{PSP}:][]{Fox2016} is equipped with a heliospheric imager that records the brightness of the corona from a vantage point situated in the corona. The Wide-Field Imager for Solar PRobe \citep[WISPR:][]{Vourlidas2016SSRv} is mounted on the ram side of the spacecraft, so the solar wind structures can be imaged prior to their in-situ measurement. According to Thomson scattering theory, as an imager gets closer to the Sun, it becomes sensitive to plasma located over a more narrow region of the solar atmosphere, acting as a microscope scrutinizing the fine-structure coronal rays compared with near 1~au based instruments \citep[][]{Vourlidas2016SSRv}. The purpose of this paper is to exploit the WISPR images from the first \textit{PSP} perihelion to gain new insights on coronal rays and on the mechanisms that form these structures and the slow solar wind.\\

The paper is structured as follows. In the first section, we describe the WISPR observations of coronal rays and their variability. In the second section, we present a technique to visualize these observations as heliographic latitude versus time (or longitude) maps. We then employ our coronal and solar wind models to simulate WISPR-like images, highlighting the successes and difficulties of the models. In the last part, we exploit the 3D nature of the modeling to interpret the features observed by WISPR.\\

\begin{figure}[t!]
%\plotone{FOV.png}
\centering
\includegraphics[scale=0.35]{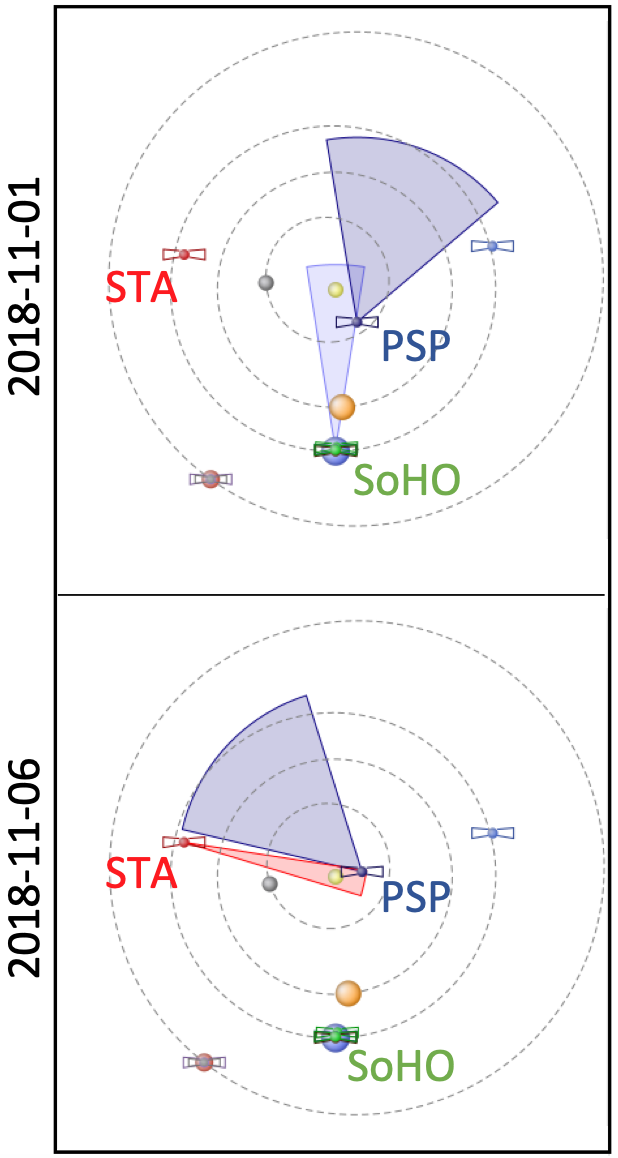}
\caption{Views of the ecliptic plane from solar north showing the relative orbital positions of STA (red), \textit{SOHO} (green) and \textit{PSP} (blue) in inertial coordinates (Heliocentric Aries Ecliptic) on 2018-11-01 (top) and 2018-11-06 (bottom). The STA COR-2A, \textit{SOHO} C3, and \textit{PSP} WISPR-I fields of view are shown with color shaded areas. The planets (Earth, Venus and Mercury) are shown as colored disks. This figure was produced with the Propagation Tool described in \citet{Rouillard2017}. \label{fig:3DPos&FOV}}
\end{figure}

\section{The WISPR imaging suite} \label{sec:WISPRImaging}

The WISPR field of view (FOV) is centered $10^\circ$ below the ecliptic plane, and it is radially offset from the Sun. The combined fields of view of the two WISPR telescopes cover a range of elongation angles (azimuthal angle away from the Sun) from 13.5$^\circ$ to 108$^\circ$ with a spatial resolution of 6.4 arcmin \citep[see][]{Vourlidas2016SSRv}. The inner (WISPR-I) telescope extends in elongation angles from 13.5$^\circ$ to 53$^\circ$ and the outer telescope (WISPR-O) extends from 50$^\circ$ to 108$^\circ$. During the first encounter, WISPR instruments obtained images from 2018 October 1 to November 10 \citep{Howard2019Nat}.  \\

Figure~\ref{fig:3DPos&FOV} presents a view of the ecliptic from solar north with the relative positions of STA, \textit{PSP}, and the planets of the inner heliosphere. It also shows the relative FOVs of the WISPR-I, \textit{SOHO} LASCO C3 \citep{Brueckner1995}, and  SECCHI/COR-2A \citep[][]{Howard2008} instruments. At the start of the encounter, on 2018 November 1, \textit{PSP} was imaging a similar part of the corona to LASCO C3, off the west limb of the Sun as viewed from Earth. In contrast, near perihelion on 2018 November 6, \textit{PSP} was imaging plasma seen also by the COR-2A instrument. This provides a great opportunity to compare the structure of the rays observed by \textit{PSP} with those imaged by LASCO and COR-2A. \\

The first perihelion was at a heliocentric distance of 35.7 \textit{R}$_\odot$. The impact parameter, the point of closest approach to the Sun along an LOS, was 8.3 \textit{R}$_\odot$ for an LOS situated at 13.5$^\circ$ the inner edge of WISPR-I. WISPR imaged in detail streamers at high cadence and high resolution inside the estimated Alfv\'en zone, which is expected to lie between 10 and 30 \textit{R}$_\odot$ \citep[e.g.][and references therein]{Sheeley2002ApJ,DeForest2014ApJ}. This is the first time from such a close distance to the Sun, inside 0.25 au. It was therefore imaging the formation of the slow solar wind that typically accelerates gradually to 30 \textit{R}$_\odot$ \citep[][]{Sheeley1997ApJ,Sanchez2017ApJ}. \\

\begin{figure}[h!]
\centering
\includegraphics[scale=0.5]{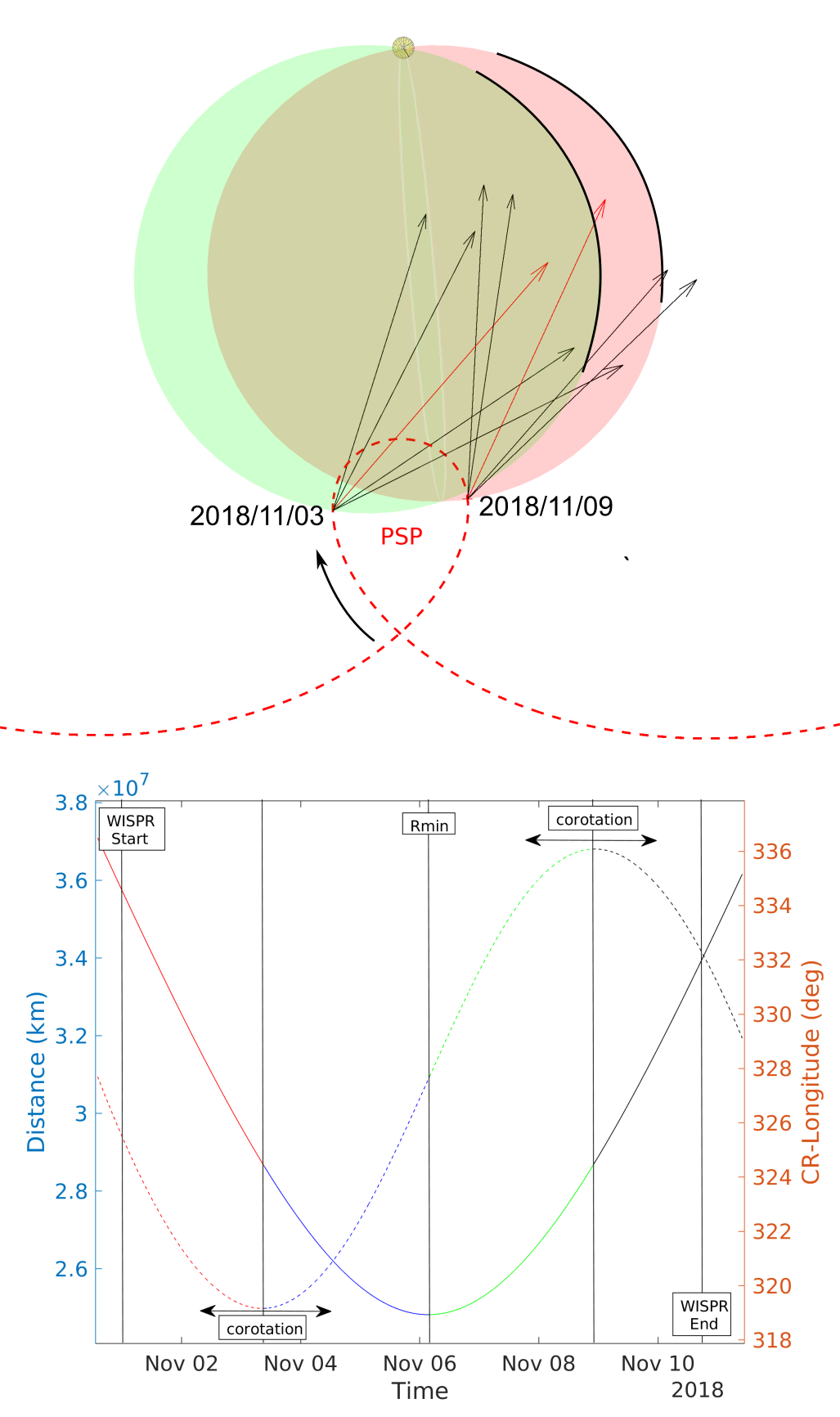}
\caption{Top panel: a view of the ecliptic from above in the Carrington coordinate system. The \textit{PSP} orbit is plotted with a red dashed line. The regions observed by WISPR-I are shown for 2018 November 3 and 9. The field of view of WISPR-I is also shown, with the black and red arrows to denote the corners and the central line of sight, respectively. The Sun is plotted to scale. Bottom panel: the heliocentric radial distance (solid line) and Carrington longitude (dashed line) of \textit{PSP} vs. time during the first encounter. The perihelion and also the two periods when \textit{PSP} was nearly corotating are labeled.  \label{fig:OrbCarThomson}}
\end{figure}

The light recorded by WISPR-I is a combination of photospheric photons either scattered by dust particles (F-corona) or electrons (K-corona). For our analysis, we used Level-3 WISPR images where the F-corona was removed by using an adapted technique developed by \citet{Stenborg2018}. In addition, the  Level-3 WISPR images were corrected for the exposure time and vignetting effects of the detector. A stellar-photometry-based technique adapted from \citet{Bewsher2012} and \citet{Tappin2015} has further improved the initial vignetting function of the instrument that was determined during preflight calibration. Each WISPR-I image is made of several (eight) short exposures $\sim$20~s which are then aggregated onboard \textit{PSP}. From 2018 November 5 to 6, WISPR-I recorded images at a higher cadence of $\sim$8 minutes compared to $\sim$45 minutes during the rest of the encounter.\\

\section{WISPR observations of coronal rays} \label{sec:ObsWISPRrays}

The brightness in WISPR-I images originates from light mostly scattered by electrons situated close to a surface called the ``Thomson sphere'' and it is located ahead of \textit{PSP}. The ``Thomson sphere'' is geometrically defined as the locus of points where the scattering efficiency is maximum according to the Thomson scattering theory \citep{Vourlidas2006ApJ}. Therefore, WISPR-I images should be much more sensitive to light originating from density structures situated along a limited portion of the LOS compared to images taken from near 1~au (see discussion in Section~\ref{sec:LASCO_imaging}). Any streamer ray that is situated close to the `Thomson sphere'' is therefore imaged in great detail and less affected by other rays \citep{Vourlidas2006ApJ}. \\

\begin{figure*}[ht!]
\includegraphics[scale=0.98]{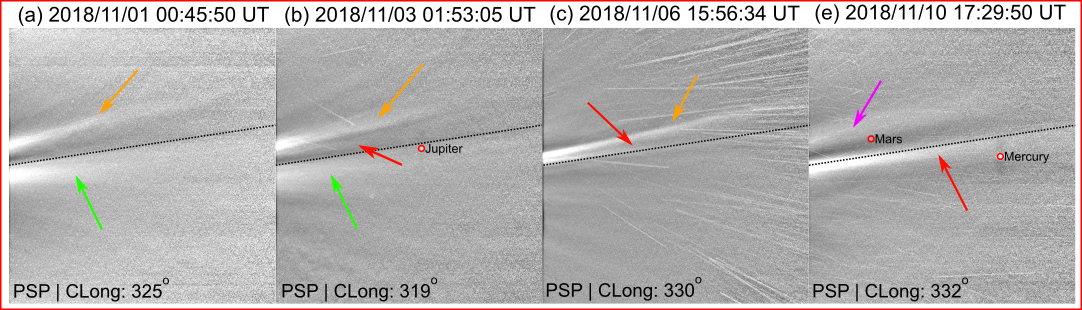}
\caption{A sequence of WISPR-I level-3 images during the first encounter from 2018 November 1, 00:45~UT to 2018 November 10, 17:29~UT. Features of interest that are discussed with more detail in the text are indicated by the colored arrows. Their color scheme will be reused consistently in the following figures and discussions. The black dashed lines mark the solar equator as reference.  \label{fig:SeqWISPR}}
\end{figure*}

The top panel of Figure~\ref{fig:OrbCarThomson} presents another view of the ecliptic with the position of \textit{PSP} in Carrington coordinates. The Thomson spheres are shown for two orbital positions, green for before and red for after perihelion. The coronal region where electrons are expected to contribute most to the visible light recorded by WISPR-I is shown as arcs on these spheres, which corresponds to the intersection between the Thomson sphere and the FOV of the instrument. During this first encounter, WISPR-I observed plasma outflows originating from a narrow range of Carrington longitudes and for an extended period. This provided a unique opportunity to disentangle the spatial from the time-dependent brightness variations of coronal rays. The Thomson sphere varies in size with heliospheric distance of the observer. We will illustrate in this paper the impact on the aspect of coronal rays of the $\sim20$ \textit{R}$_\odot$ change in heliocentric radial distance executed by \textit{PSP} between the start of the encounter (2018 November 1) and perihelion (2018 November 6). This effect will likely be even more critical for the interpretation of future WISPR observations. \\

The fast spacecraft motion in the sense of solar rotation was such that \textit{PSP} remained in quasi-corotation and even super-rotation with the low-latitude corona during the encounter. This can be seen in Figure~\ref{fig:OrbCarThomson}, bottom panel, where we show the evolution of \textit{PSP}'s radial distance and Carrington longitude during the encounter. The Carrington longitudes of the spacecraft only changed by 17$^\circ$, remaining between 319$^\circ$ and 336$^\circ$ between 2018 November 1 and 10. This essentially means that during the first encounter, WISPR imaged plasma flows originating from a small region of the solar corona. For comparison, the Earth's Carrington longitude changed by over 132$^\circ$ during that same time interval. Figure~\ref{fig:OrbCarThomson}, bottom panel, also shows the different phases of spacecraft motion. Between 2018 November 4 and 8, \textit{PSP} motion was in super-rotation with the low-latitude corona; outside this interval it was either in quasi-corotation or under-rotation. One last aspect of the evolution of \textit{PSP} heliolongitude of Figure~\ref{fig:OrbCarThomson}, bottom panel, is that WISPR imaged different parts of the solar corona before and after perihelion. \\ 

Figure~\ref{fig:SeqWISPR} presents a sequence of images from WISPR-I during the first encounter at times when no CME was passing in the FOV. In this work, we focus on the analysis of images from the WISPR-I telescope only, as its inner FOV provides the finest and clearest observations of coronal rays. The images show the presence of multiple rays whose position and brightness evolve significantly during the encounter. Specific rays of interest are annotated by colored arrows to guide the reader through section~\ref{sec:NumModels} and section~\ref{sec:Interpretation}. \\

A bright ray (orange arrows) located just a few degrees north of the equatorial plane is visible during most of the encounter. From 2018 November 3 (panel b), this coronal ray seems to split with an additional ray (red arrows) appearing in the image. A diffuse and fainter ray (green arrows) is also observed south of the equatorial plane during the first half of the encounter (panels a and b), which then disappears. The northern and southern rays are separated by a thick dark band. This dark feature is likely induced by the background removal of the F-corona and may disappear in future versions of the level-3 WISPR images. At the end of the encounter (panel e), WISPR-I scans a new coronal region (see Figure~\ref{fig:OrbCarThomson}) as a new broad diffuse ray (magenta arrow) emerges from northern latitudes. \\ 

\section{Comparing with 1~au imaging (SECCHI and LASCO)} \label{sec:LASCO_imaging}

 Alignments of WISPR with LASCO C3 and SECCHI COR-2A FOVs provide an opportunity to compare the same structures imaged by different spacecraft. Figure~\ref{fig:3DPos&FOV} shows that there are two periods when WISPR imaging is directly comparable with SECCHI and LASCO images. At the start of the WISPR science observing window (2018 November 1), LASCO C3 imaged similar plasma above the west solar limb of the Sun as viewed from Earth. The second opportunity was on 2018 November 6, when WISPR-I and COR-2A were imaging plasma from the same region above the eastern solar limb as viewed from STA. \\

Comparing images from different instruments can be a challenging task, especially from separate observatories. In addition, accounting for the match of the Thomson spheres between instruments makes this comparison even more complex. For example, one can see in the \href{https://nuage.irap.omp.eu/index.php/s/x2tHzaBM9fyL1bt}{supplementary movie} associated with Figure~\ref{fig:frames_movie} in the Appendix that the best match between the \textit{PSP} and \textit{SOHO} Thomson spheres occurred $\sim$1 day before the start of the WISPR observing window. In the previous section, we reinforced the importance of the Thomson sphere as the surface where the contribution to the total brightness along each LOS is expected to be maximum. But in practice, there are also noticeable contributions away from this surface and hence a broader region must be considered. This is particularly true for observations taken from near 1~au. Figure~\ref{fig:Thomson_fig}, given in the Appendix, shows a comparison of the regions likely observed by WISPR-I and LASCO C3 on 2018 November 1 as predicted by the Thomson scattering theory. It shows that WISPR-I and LASCO C3 imaged a similar coronal region. As we can see in Figure~\ref{fig:Thomson_fig}, the brightness recorded by WISPR-I originates from a broad region along the LOS that extends from \textit{PSP} to well behind the Thomson sphere. As \textit{PSP} gets closer to the Sun, this region shrinks. Nevertheless, Figure~\ref{fig:Thomson_fig} shows that WISPR-I already records plasma brightness from a smaller region than LASCO C3. \\

\begin{figure*}[ht!]
\centering
\includegraphics[scale=0.8]{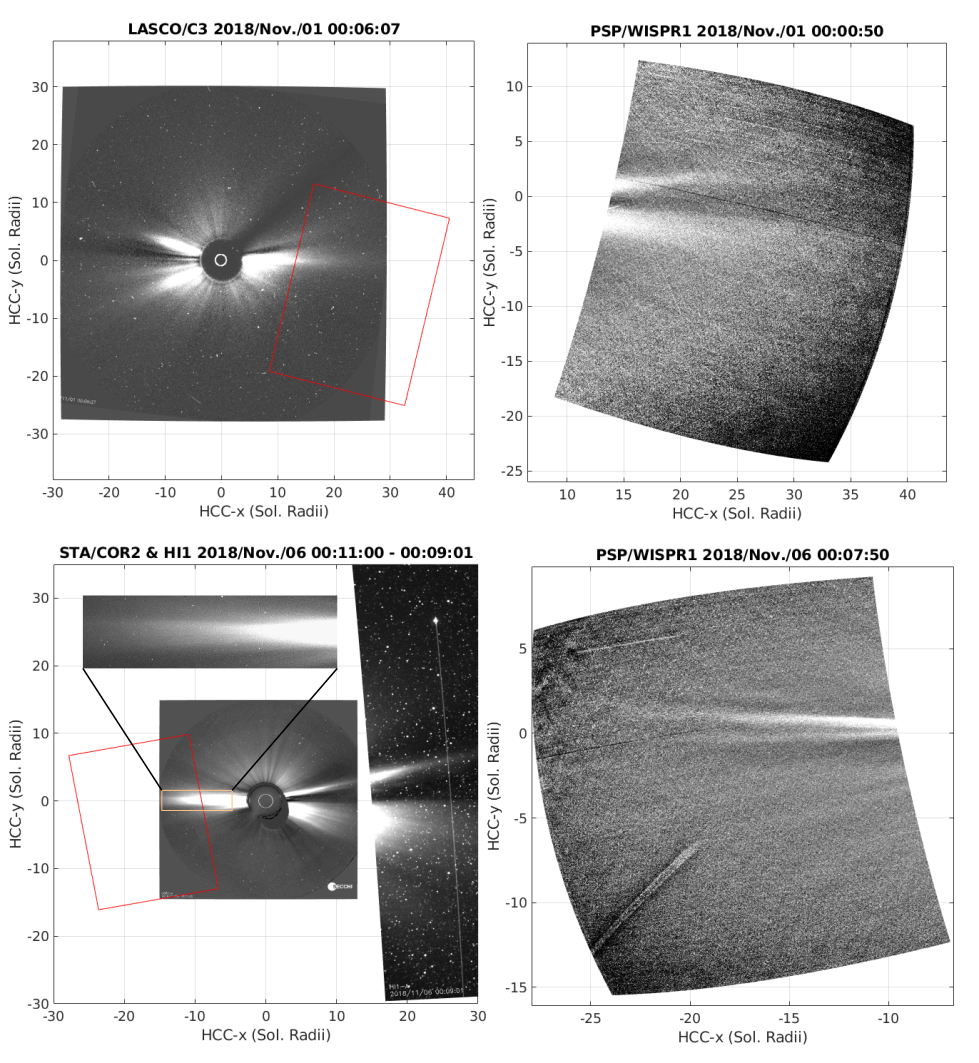}
\caption{A comparison between WISPR-I images with LASCO C3 images at around 00:06 UT on 2018 November 1 (top panels) and STA COR-2A and HI1 images at around 00:11 UT on 2018 November 6 (bottom panels). The red boxes overplotted at the left panels represent the WISPR-I field of view for comparison with those observations. \label{fig:CompWLImagersPlain}}
\end{figure*}

Considering the above complications, we give in Figure~\ref{fig:CompWLImagersPlain} an overview of the zoomed-in view offered by WISPR-I on the typical streamer structures observed from near 1~au observatories. In Figure~\ref{fig:CompWLImagersPlain}, we compare WISPR-I with LASCO C3 and COR-2A observations for two selected days when the comparison was optimal. On 2018 November 1 (top panels), the two brightest rays situated just a few degrees north and south (orange and green arrows in Figure~\ref{fig:SeqWISPR}(a)) of the equatorial plane are imaged by both spacecraft. In addition to these two bright features, a number of much fainter rays are also visible at \textit{PSP}, unveiling an apparent complex structuring of the corona, which is not resolvable in LASCO C3 images. Besides, we note that, on 2018 November 1, WISPR observations had not reached their highest resolution for this encounter yet, as \textit{PSP} was located at 51~\textit{R}$_\odot$, e.g., $\sim$16~\textit{R}$_\odot$ away from perihelion. \\

On 2018 November 6, the best alignment was with STA, and Figure~\ref{fig:CompWLImagersPlain} (bottom panels) shows a comparison of COR-2A and WISPR-I images. The pair of bright rays (orange and red arrows in Figure~\ref{fig:SeqWISPR}(c)) located just above the equatorial plane is not clearly distinguishable in COR-2A despite the adequate resolution of the instrument to resolve such structures. This is likely an LOS integration effect, suggesting that 1~au may be not capturing the fine coronal structure. The zoomed-in view that WISPR offers and the shrinking of the Thomson sphere, predicted by \citet{Vourlidas2016SSRv}, make it possible to observe coronal rays in finer detail compared with near 1~au observations. \\

\section{WISPR Carrington maps} \label{sec:CARR_WISPR}

The representation of WL imagery in the form of heliocentric latitude versus longitude maps, such as ``Carrington maps'', provides a powerful tool to interpret the time and spatial evolution of coronal structures observed in a sequence of images (see, for example, the review by \citep{Rouillard2019}). These maps, built from coronagraph images, have been exploited in many studies on the origin of real streamers and pseudostreamers \citep[e.g.][]{Wang2007ApJ}. These maps are built by assuming that the brightness of a pixel results from a plasma parcel located at a point P along the LOS where the impact parameter is minimized. This is precisely the locus of points P at the Thomson sphere \citep{Vourlidas2006ApJ,Howard2012ApJ}. In essence, the Carrington maps are constructed by first calculating the elongation and position angle of LOS that intersect the Thomson sphere at the heliocentric radial distance of interest. The associated pixels are then mapped in a Carrington longitude versus latitude format, and this procedure is repeated for each image.  \\

\begin{figure}[ht!]
\centering
\includegraphics[scale=0.23]{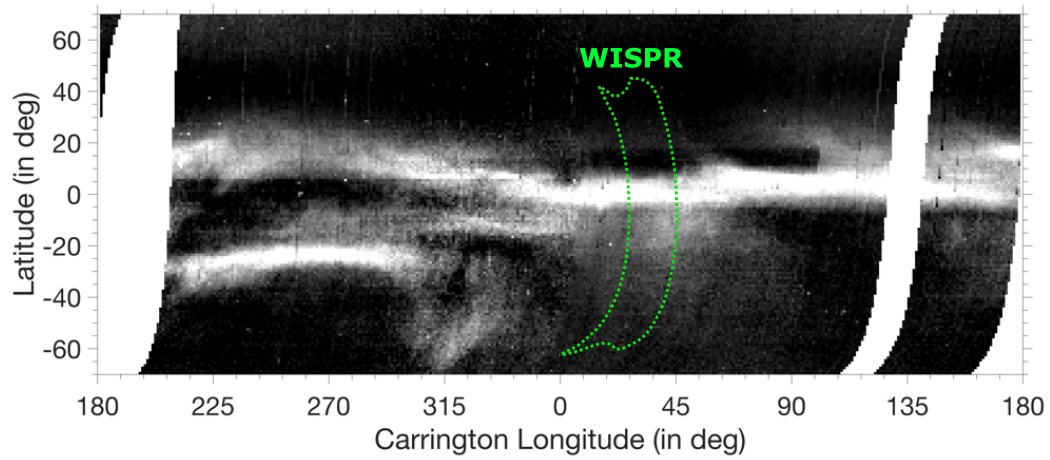}
\caption{Carrington map from LASCO C3 white-light observations during the time interval 2018 October 20 to November 14. 
\label{fig:LASCO_map}}
\end{figure}

Such a map is shown in Figure~\ref{fig:LASCO_map} based on LASCO C3 observations during the \textit{PSP} first encounter. The extent in Carrington latitude and longitude of the same map based on WISPR-I observations is superimposed in green; this emphasizes the zoomed-in view of WISPR-I compared with LASCO observations. A \href{https://nuage.irap.omp.eu/index.php/s/x2tHzaBM9fyL1bt}{supplementary movie} (Figure~\ref{fig:frames_movie}) shows how the WISPR-I Carrington map is built over time. The WISPR-I Carrington map only covers $\sim$30$^\circ$ of Carrington longitudes (as seen in Figure~\ref{fig:OrbCarThomson} as well). One can also notice that due to the super-rotation phase, WISPR-I imaged the same coronal region twice. This is also illustrated in Figure~\ref{fig:OrbCarThomson} by the loop shape of the \textit{PSP} orbit in the Carrington frame. This makes the Carrington map format somewhat cumbersome to interpret, as shown in the \href{https://nuage.irap.omp.eu/index.php/s/x2tHzaBM9fyL1bt}{supplementary movie}. \\

\begin{figure*}[ht!]
\centering
\includegraphics[scale=0.8]{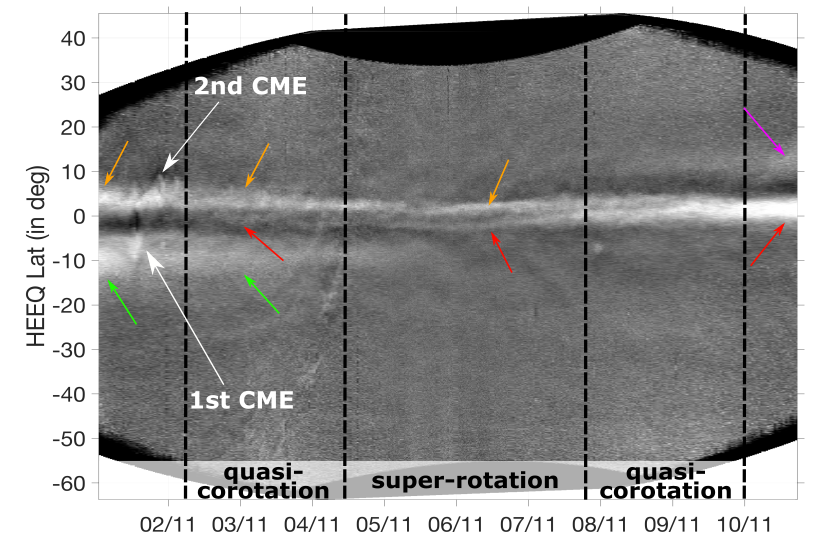}
\caption{Latitude vs. time map from WISPR-I images taken during the first encounter: from 2018 November 1, 00:45~UT, to 2018 November 10, 17:29~UT. The y-axis is the heliographic latitude in degrees.  
\label{fig:WISPR_maps}}
\end{figure*}

Hence, it is more convenient to build a map in a coordinate system that is not rotating with the Sun. Indeed, similar maps can be constructed by considering instead the Heliocentric Earth Equatorial (HEEQ) coordinate system; we refer to such a map here as a HEEQ-map. Alternatively, because each image has an associated observation time, we can also build a map in a latitude versus time format. Examples of such maps built with \textit{STEREO} heliospheric images were exploited in \cite{Rouillard2010ApJ} to analyze the source regions of the slow wind and the formation of corotating interaction regions. Figure~\ref{fig:WISPR_maps} presents the WISPR-I latitude versus time format for the first encounter. Because this format is easier to interpret, we focus our analysis on this map in the remainder of this paper. \\

From the map of Figure~\ref{fig:WISPR_maps}, we recognize a number of interesting features seen in Figure~\ref{fig:SeqWISPR} (see Section~\ref{sec:ObsWISPRrays}), and they are annotated with the same colors. The two bright and thick horizontal stripes a few degrees north (orange arrow) and south (green arrow) of the equatorial plane from 2018 November 1st to 5th correspond to the two bright rays that were identified in both WISPR and LASCO C3 images on November 1st (see Figure~\ref{fig:SeqWISPR} and Figure~\ref{fig:CompWLImagersPlain}). The southern streamer ray progressively becomes dimmer and after 2018 November 6 is almost untraceable. A reason why this ray does not clearly reappear later on in the map could be that on 2018 November 10, \textit{PSP} has not come back yet to its initial position of 2018 November 1 (see Figure~\ref{fig:OrbCarThomson}). Furthermore, we mentioned in Section~\ref{sec:ObsWISPRrays} the presence of a dark feature between the northern and southern rays that is likely due to the background model of the F-corona which is subtracted from the WISPR images. This dark feature is clearly visible in the map shown in Figure~\ref{fig:WISPR_maps} and might make the southern streamer ray darker than it should be after 2018 November 6. Another dark band is also visible above the northern streamer ray after 2018 November 7, which may also be a side effect of the background removal.  \\

The northern streamer appears to be more variable than the southern one. In Section~\ref{sec:NumModels} we will show that the southern rays emerge from a pseudo-streamer. At this period, \textit{PSP} is nearly co-rotating and WISPR is imaging almost the same coronal region hence the variability at the northern streamer is mostly produced by the time-dependent effects associated with the release of small-scale transients. We observe some brightening located at the core of the northern streamer and some bright spike-like features that appear periodically near the streamer edges. These are transient perturbations of the streamer associated with the passage of two consecutive CMEs \citep{Hess2020ApJ,Rouillard2020ApJ} indicated by two white arrows on the WISPR map. The southern streamer is more diffuse and exhibits much less activity with the exception of the passage of the first CME which originated in the northern streamer and was deflected southward during its transit through the corona to WISPR. This deflection discussed by \citet{Hess2020ApJ} led to the disruption of the pseudostreamer situated on the southern flank of the CME. \\

Despite the narrow range of Carrington longitudes imaged by WISPR, the different rays change significantly over the time interval considered here. This is most pronounced during super-rotation (from 2018 November 4 to 8) when the spacecraft starts imaging plasma from different longitudes. The northern streamer (orange arrows) become thinner and dimmer, and substructuring progressively appears where the streamer splits into two lanes (red arrows). Some traces of this splitting can be observed between 2018 November 4 and 5 but it is best observed after 2018 November 5 and until 2018 November 9. The dark band separating the two bright streamer lanes (orange and red arrows) remains near the equatorial plane and then shifts progressively a few degrees northwards until 2018 November 9. This transition occurs when WISPR is expected to observe a new coronal region (see Figure~\ref{fig:OrbCarThomson}). \\

After 2018 November 9, the splitting of the rays (orange and red arrows) progressively disappears and a single ray forms again (indicated by a magenta arrow) at around 10$^\circ$ north. This new ray is dimmer and diffuse without any particular structuring or activity. As we can see already in Figure~\ref{fig:OrbCarThomson}, WISPR was by then imaging a different part of the corona than at perihelion.\\

Thanks to the proximity of \textit{PSP} to the Sun, WISPR is sensitive to a limited region along the LOS. As shown in Figure~\ref{fig:WISPR_maps}, WISPR is able to capture very faint small irregularities even in the thinnest point of the streamer rays at closest approach on 2018 November 6. The thickness of the northern ray is estimated to be $\approx$5$^\circ$, and this is similar to the typical thickness of the HPS measured in situ by spacecraft crossing \citep[e.g.,][]{Winterhalter1994JGR}. The HPS originates from the densest part of the streamers and usually contains many density substructures \citep[][]{Sanchez2017ApJ}. The small irregularities visible in the WISPR-I images might be associated with such density substructures. As \textit{PSP} gets closer to the Sun on its way to 10~\textit{R}$_\odot$, WISPR will be likely able to further resolve the fine-scale structures of the HPS. \\

Some other intriguing faint features appeared during super-rotation near the northern edges of the upper streamer lane. They emanate from the bright northern ray and drift to higher northern latitudes. Similar features occur near the southern edges of the streamer lane but they are fainter and some appear in isolation. All these drifting structures cease to occur on 2018 November 9. In order to make sense of these WISPR-I observations, we now exploit in the next sections high-resolution simulations of the solar corona and of the escaping solar wind. \\ 

\section{Comparison with numerical models} \label{sec:NumModels}

In this section, we introduce a 3D magnetohydrodynamic (MHD) model of the solar wind that we exploit later on in section \ref{sec:Interpretation} to study the origin of the coronal structures observed by WISPR discussed above. The purpose of the modeling is to provide a supporting role to interpret the observations and also to illustrate the various difficulties faced in the interpretation of WISPR data. \\

We simulate the corona and solar wind using the 3D multitube MHD code called MULTI-VP \citep[see][]{Pinto2017ApJ}. MULTI-VP computes the properties of the solar wind such as speed, density, and temperature by solving a set of 1D (MHD) conservation equations along individual flux tubes. The energy equation includes the effect of heating, thermal conduction, and radiative cooling, which are essential in order to simulate a realistic solar wind mass flux \citep{Hansteen1995JGR,Pinto2017ApJ}. This model has been run on thousands of magnetic flux tubes to simulate the entire solar wind escaping the solar atmosphere. The inner boundary of the simulation domain is at the photosphere and extends typically to about 30~\textit{R}$_\odot$. For the purpose of our study, the outer boundary of the MULTI-VP simulation was set to 90~\textit{R}$_\odot$ in order to include the brightness contribution of electrons situated far behind the Thomson sphere.  \\

The MULTI-VP solar wind model runs on coronal magnetic fields that can be derived by potential field source-surface (PFSS) extrapolations of magnetograms measured by different observatories. In this study, we used photospheric magnetic field maps from the Wilcox Solar Observatory (WSO) and those computed by the Air Force Data Assimilative Photospheric Flux Transport model (ADAPT). Hereafter we will refer to these two sets of simulations as MVP-WSO and MVP-ADAPT. The ADAPT maps are constructed with GONG magnetograms, and a flux transport model is used to simulate the motion of photospheric magnetic fields \citep[see][]{Arge2010}. The maps also exploit ensemble least-squares (EnLS) data assimilation method that account for model and observational uncertainties  \citep{Worden2000SoPh}. \\

The MULTI-VP simulations have a grid adapted to the input magnetogram resolution. The MVP-WSO run has a grid of 5$^\circ$ resolution in both latitude and longitude, and the MVP-ADAPT run is at higher grid resolution of 2$^\circ$. The MVP-ADAPT simulation provides 2.5 times finer resolution compared with the MVP-WSO run; this provides a significant impact on the synthetic images which will be discussed later in this section. We also note that the ADAPT input magnetogram which employs a flux transport model is expected to give more realistic polar magnetic fields. This could affect the latitudinal extent of the coronal neutral line and in turn the position of streamer rays. The main objective of the study is to exploit the modeling to interpret the first observations made by WISPR-I and in particular the evolution of the large-scale structures. \\

\begin{figure*}[t!]
\centering
\includegraphics[scale=0.4]{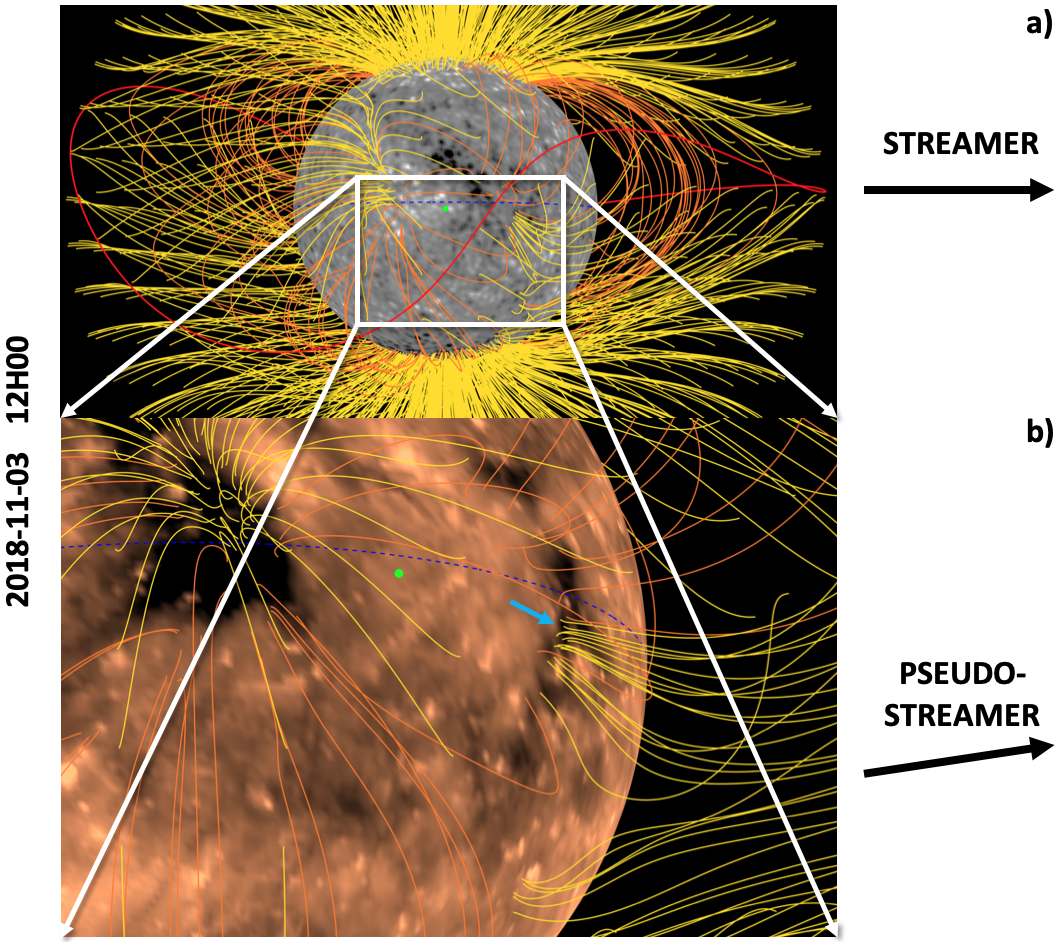}
\caption{Panel (a): a 3D view of the magnetic field lines reconstructed using the PFSS method based on an ADAPT magnetogram (displayed in gray scale) produced for 2018 November 5 at 12:00~UT. The viewpoint is placed at the position of \textit{PSP} on 2018 November 3 at 12:00~UT. The open and closed magnetic field lines are depicted in yellow and orange, respectively. The polarity inversion line (neutral line) at the source surface is plotted as a red line. Panel (b): a zoomed-in view of the pseudostreamer that entered the field of view of WISPR-I. Color plotted is a Carrington map in the 193~$A^\circ$ Extreme UltraViolet (EUV) wavelength, from the combined vantage points offered by STA and the \textit{Solar Dynamics Observatory} on 2018 October 30 at 12:00~UT. \label{fig:PFSSextrap}}
\end{figure*}

In Figure~\ref{fig:PFSSextrap}(a), we present the results of a PFSS extrapolation based on the ADAPT magnetogram produced for 2018 November 5 at 12:00~UT. The source surface was set at a height of 2.5 \textit{R}$_\odot$ for this calculation. The viewpoint was set at \textit{PSP} on 2018 November 3 at 12:00~UT. WISPR-I was observing plasma that originated near the west limb of the Sun (e.g. on the right-hand side of the image). The open and closed magnetic field lines are shown in yellow and orange, respectively, and the position of the coronal neutral line, i.e. the origin of the HCS, at the source surface is given by the red line. The cusp of a helmet is identified near the origin of the brightest rays imaged by WISPR off the west limb just a few degrees north of the equator. This streamer is east-west oriented in that region and produces the narrow (in latitude) band of rays. We now refer to these coronal rays (marked with the orange arrows) as ``streamer rays''. The above results are also valid for the extrapolations based on WSO. \\

Figure~\ref{fig:PFSSextrap}(b) presents a zoomed-in view of panel (a) centered to the south of the streamer. From 2018 November 1 to 10, this region remains backsided from STA and Earth-based observatories, thus a past EUV map on 2018 October 30 is shown, which covers this region. The coronal hole (indicated by a blue arrow) is assumed to not have significantly evolved until the time of the PFSS extrapolation on 2018 November 5. We identify the presence of a bundle of magnetic field lines rooted in a small region located between the position of \textit{PSP} and the west limb. The EUV map reveals the presence of an isolated coronal hole (blue arrow) near the footpoint of these field lines with negative polarity. The magnetic field lines from that region meet open magnetic field lines of the same polarity that are rooted in the southern coronal hole to form a low-lying cusp-like structure. This is a typical pseudostreamer, and it is located approximately 20$^\circ$ south of the equator and west of central meridian as viewed from \textit{PSP}. The coronal rays located below the equatorial plane (green arrow) and observed by WISPR-I (see green arrows in Figure~\ref{fig:SeqWISPR}) are likely associated with this pseudostreamer. These rays are located off the plane of the sky toward the observer.  \\

Above the source surface, here placed at 2.5 \textit{R}$_\odot$, MULTI-VP provides densities in all regions of space occupied by open magnetic fields. Using those cubes, we produce synthetic WL imagery by applying the theory of Thomson scattering \citep[see][]{Howard2009Ssr}. These synthetic images are compared with WISPR observations, providing a baseline to interpret the imaged structures. For the construction of the synthetic WL images we use the following procedure. For each WISPR-I image, we calculate the heliographic coordinates of all pixels situated at the Thomson sphere. We define LOSs that start from \textit{PSP}, pass to those points, and extend beyond the Thomson surface. Along the LOS, we interpolate the solar wind density values from the MULTI-VP model, and we use the Thomson scattering theory to calculate an intensity for each point defined along the LOS \citep[see][]{Howard2009Ssr}. We integrate the brightness values along each LOS to retrieve the total brightness of each pixel. \\

\begin{figure*}[ht!]
\centering
\includegraphics[scale=0.65]{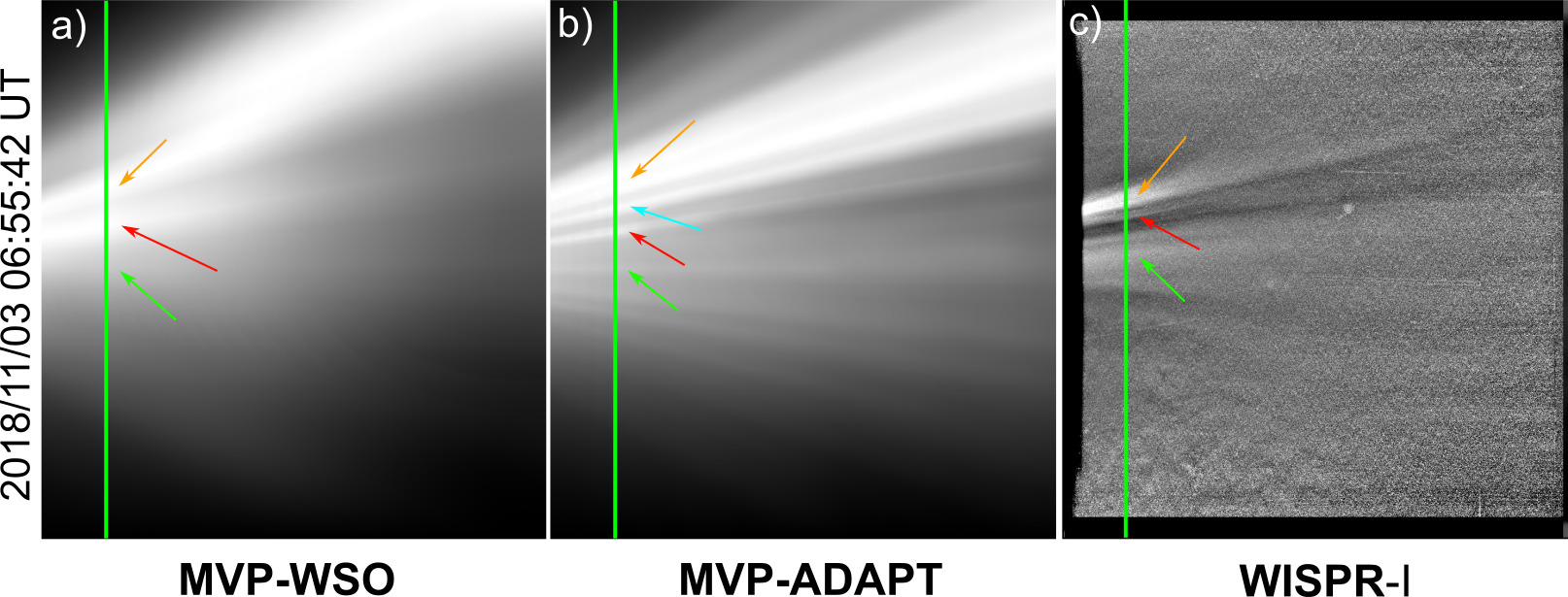}
\caption{ A comparison between synthetic WL images (panels a and b) and a WISPR image (panel c) on 2018 November 3, at 06:55~UT. The synthetic images produced by MHD data from the MULTI-VP model and two different source magnetograms; WSO magnetogram of Carrington rotation CR2210 (panel a) and ADAPT magnetogram of 2018 November 5, 12:00~UT (panel b). The arrows superimposed at the images are the same as in Figures~\ref{fig:SeqWISPR} and~\ref{fig:WISPR_maps}, and are indicative of the position of the features discussed in the text. \label{fig:WLSynthISPR}}
\end{figure*}

Figure~\ref{fig:WLSynthISPR} presents examples of synthetic WL images produced from the MVP-WSO (panel a) and MVP-ADAPT (panel b) runs. Panel (c) shows the corresponding WISPR-I image on 2018 November 3 at 06:55~UT for comparison with the simulated images. For illustration purposes, we have rescaled the intensity for both synthetic WL images to enhance the visibility of the streamer rays. As a consequence, we perform a qualitative comparison between synthetic products and real observations in the following sections without considering differences on the features intensity in WL. However, for completeness, we provide in Figure~\ref{fig:vert_slice} the distribution of the normalized intensity along vertical slices (green lines) that gives a comparison of the coronal-rays latitude in the synthetic and WISPR-I images.\\

Comparing the synthetic images between the two runs, we find, as explained earlier, similarities in the features observed near the equator, but there are also some striking differences that will be presented later on in the text. Overall, the high-resolution MVP-ADAPT simulation produces a more detailed view of the streamers and reveals substructures that are absent in the MVP-WSO simulation. The streamer rays (orange arrows) are reproduced by both MVP-WSO and MVP-ADAPT. This is also the case for the pseudostreamer identified in Figure~\ref{fig:PFSSextrap}(b) and here pointed with the green arrows.\\

The splitting of the northern streamer into two lanes observed in the WISPR-I image (orange and red arrows in panel c) and discussed in Section~\ref{sec:ObsWISPRrays} is also well reproduced in both simulations but it appears at a slightly different latitude compared to the observations (see Figure~\ref{fig:vert_slice}). As discussed earlier on in this section, it is probably related to the inherent uncertainties in polar field measurement. At first sight, it can be hard to identify the two-lane splitting within the multiple adjacent thin rays visible in the MVP-ADAPT image. Section~\ref{sec:Interpretation} will remove this ambiguity and confirm the identification made here. \\

Most of the differences between the two synthetic WL images of Figure~\ref{fig:WLSynthISPR} appear in the northern streamer, where the higher-resolution MVP-ADAPT simulation shows an additional subdivision of the northern streamer into at least two separate rays (orange and blue arrows). However, this subdivision is not discernible in the WISPR image (panel c). The more diffuse streamer rays that appear in the MVP-WSO image could result from multiple rays unresolved in this lower resolution simulation but clearly seen in the higher-resolution MVP-ADAPT run (panel b). Overall, the MVP-ADAPT simulation captures better the finer-scale structures of coronal rays that are also observed by WISPR-I in the streamer and pseudostreamer. However, the MVP-ADAPT synthetic image tends to show additional streamer rays in the northern streamer that are not seen clearly in the WISPR-I image. \\

\begin{figure*}[th!]
\centering
\includegraphics[scale=0.48]{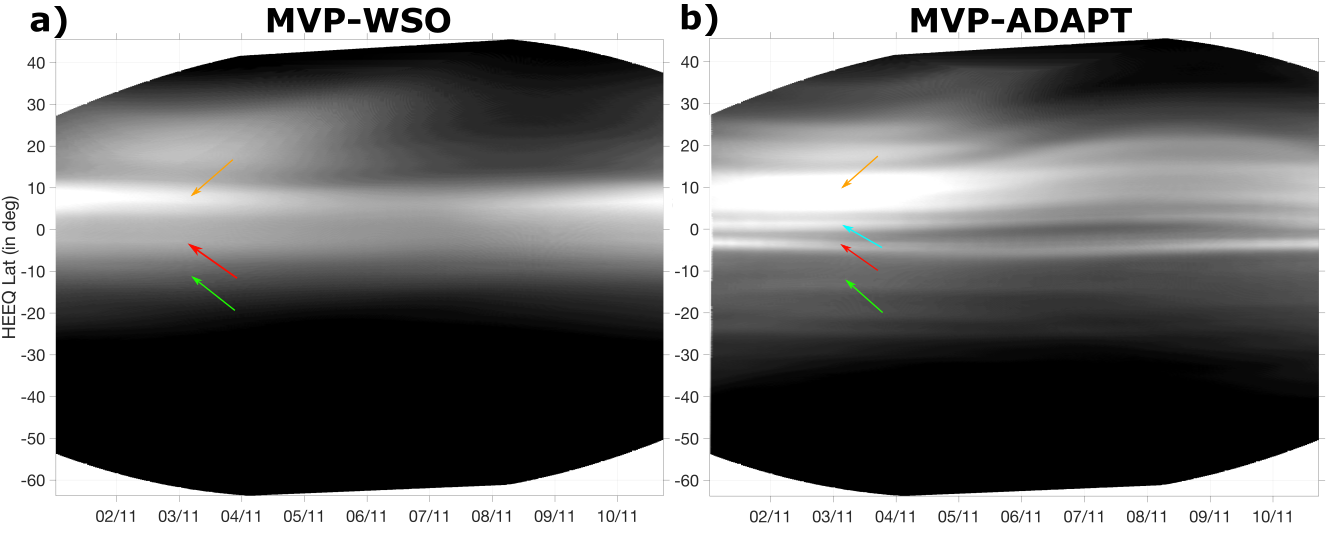}
\caption{Synthetic latitude vs. time maps processed from the WL synthetic images based on MULTI-VP simulations, in panel (a) using WSO magnetograms and in panel (b) using ADAPT modeled magnetograms. The arrows' color coding is the same as in Figures~\ref{fig:SeqWISPR},~\ref{fig:WISPR_maps}, and \ref{fig:WLSynthISPR}.  \label{fig:WLSynthCarr}}
\end{figure*}

In order to compare the hundreds of simulated and real images more clearly, we proceed by constructing synthetic WL latitude versus time maps similar to those presented in Section~\ref{sec:CARR_WISPR}. Figure~\ref{fig:WLSynthCarr} shows two synthetic WL maps produced using the same procedure presented in Section~\ref{sec:CARR_WISPR} and based on MVP-WSO (panel a) and MVP-ADAPT (panel b) density cubes. These maps are built for a heliocentric radial distance of 15~\textit{R}$_\odot$ and therefore are comparable with the WISPR map shown in Figure~\ref{fig:WISPR_maps}. Overall, the synthetic maps show a summary of the features observed in the WISPR-I images presented in Figure~\ref{fig:SeqWISPR} and Figure~\ref{fig:WISPR_maps}.\\ 

A comparison of the two synthetic maps confirms that the increased resolution of MVP-ADAPT produces fine-scale structures including the subdivision of the northern streamer (indicated by the blue arrow). Some WL rays in the MVP-ADAPT run are also visible in the MVP-WSO synthetic map but they are more diffuse. The bright streamers (orange and green arrows) appear in both maps, and the main bright structures are similar and roughly consistent with the WISPR observations shown in Figure~\ref{fig:WISPR_maps}.  \\

There are notable differences nevertheless: the two-lane splitting in the northern rays, indicated by the orange and red arrows, is visible through the whole encounter period whereas this feature is only visible in the WISPR map from 2018 November 3 to 9. Both simulations show the southern ray observed by WISPR from 2018 November 1 and 5 that originates from the pseudostreamer shown in Figure~\ref{fig:PFSSextrap}. This feature seems to disappear after 2018 November 5 in the WISPR map of Figure~\ref{fig:WISPR_maps}. On the contrary, this streamer ray is visible from the beginning until the end of the synthetic maps at both MVP-WSO and MVP-ADAPT simulations. However, as mentioned in Sections~\ref{sec:ObsWISPRrays} and ~\ref{sec:CARR_WISPR}, the southern ray dimming in the WISPR map is probably a side effect of the F-corona background removal applied to the level-3 WISPR-I images.\\

Other interesting features are mostly visible in the MVP-ADAPT map from 2018 November 5 to 8: bright rays appear above the northern streamer and seem to migrate northwards away from the streamer. These migrating structures appear in the WISPR-I map (see Figure~\ref{fig:WISPR_maps}), but they are significantly faint. Similar migrating features appear during the same period in the southern regions at the latitudes of the pseudostreamer (that has by then disappeared). These southern migrating rays are slightly visible in both the MVP-ADAPT simulations and WISPR-I observations. \\

\section{Modeling-based Interpretation of the observations}
\label{sec:Interpretation}

\begin{figure*}[ht!]
\includegraphics[scale=0.47]{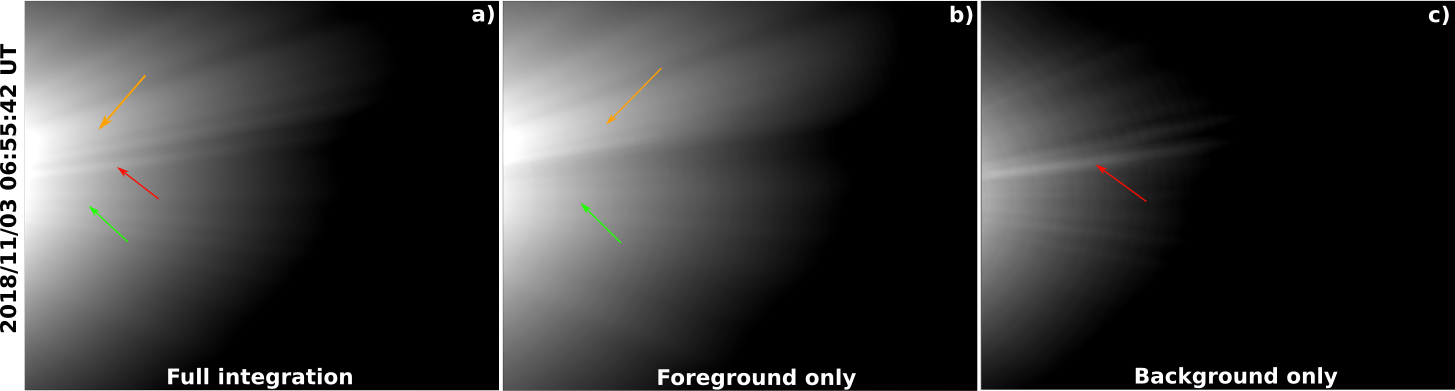}
\caption{Three WISPR WL synthetic images on 2018 November 3 at 06:55~UT from the MVP-ADAPT run. They correspond to integration along the LOS over (a) the full span, (b) the foreground only, and (c) the background only. The arrows' color coding is the same as in the previous figures. \label{fig:ADAPT_3imgs}}
\end{figure*}

Here, we further exploit the simulations to interpret the observations of the streamer and pseudostreamer rays as well as the origin and evolution of the migrating rays. The analysis of all the features appearing in the real/synthetic images and maps has proven to be a complex task. Mainly, the LOS effects make it difficult when we need to identify the source regions responsible for the different rays visible in the synthetic maps or images. \\ 
To get a better insight of where the bright rays are situated relative to the Thomson sphere, we recomputed the synthetic images by splitting the integration path along each LOS in two separate domains. The first domain covers only the region from the observer (\textit{PSP}) up to the Thomson sphere (``foreground'' region), while the second extends far out and beyond the Thomson sphere (``background'' region). In Figure~\ref{fig:ADAPT_3imgs}, the initial WL synthetic image from the MVP-ADAPT run (Figure~\ref{fig:WLSynthISPR}(b)) is again plotted in panel (a) along with its associate foreground (panel b) and background (panel c) subimages. The foreground subimage (panel b) looks very similar to the full image (panel a) and contributes to most of the diffuse brightness of the broad northern and southern rays (annotated by the orange and green arrows). In contrast, the background subimage (panel c) only reveals the thin and bright central ray at a few degrees south (marked by the red arrow), which is not visible in the foreground subimage. This is a clear hint that the full synthetic images consist of rays located over an extended region in front of and beyond the Thomson sphere. As already discussed in section \ref{sec:LASCO_imaging}, there are indeed non-negligible contributions to the total brightness on both sides of the Thomson surface that are included in the LOS integration domain. Figure~\ref{fig:Thomson_fig} in the Appendix illustrates the spatial extent of these contributions according to the Thomson scattering theory. \\

\begin{figure*}[ht!]
\begin{center}
\includegraphics[scale=0.34]{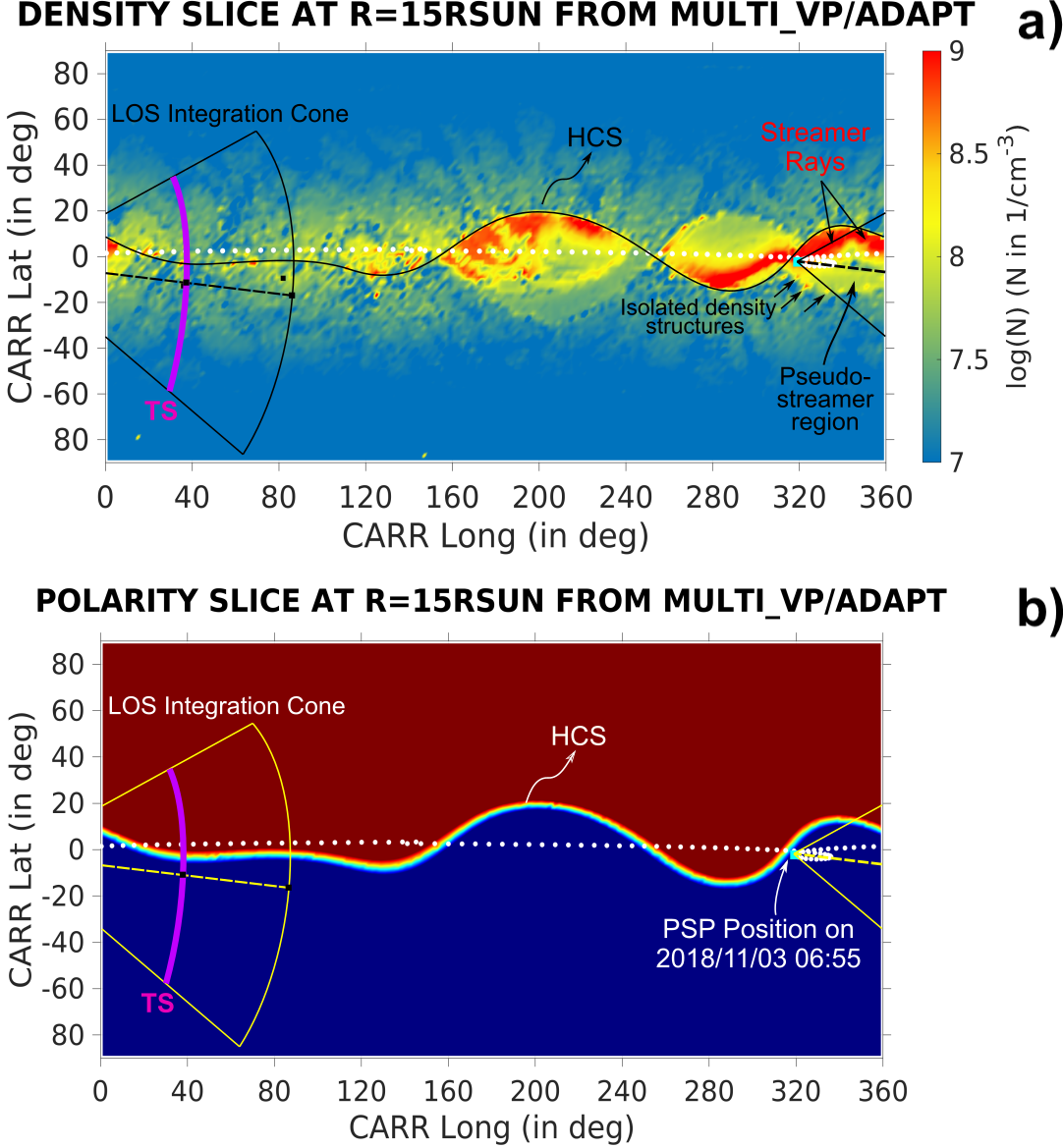}
\caption{Panels (a) and (b) show, respectively, the Carrington maps of the simulated density and magnetic field polarity at 15~\textit{R}$_\odot$ from MVP-ADAPT. The magenta line traces the intersection of the Thomson sphere with the Carrington map. The cone of integration, defined by the intersection between the field of view of WISPR-I and the map, is shown in black (yellow) in panel (a) (b). \label{fig:CarrModels}}
\end{center}
\end{figure*}

In order to understand these subimages in more detail, we investigate further the 3D topology of the corona. For that we use the density in the simulations as a proxy to visualize in 3D these bright structures. Figure~\ref{fig:CarrModels}(a) shows a Carrington map of simulated density (MVP-ADAPT) at 15~\textit{R}$_\odot$. The white dashed line represents the \textit{PSP} projected trajectory, and the green square is the \textit{PSP} position on 2018 November 3 at 06:55 UT. The magenta line separates the foreground from the background integration domain. Figure~\ref{fig:CarrModels}(b) shows the Carrington map of the magnetic polarity from the MVP-ADAPT run. Comparing both Carrington maps, we can see that the densest solar wind forms in these simulations around the HCS. Dense wind also forms along arcs that connect different parts of the HCS; these correspond to the cusp of the pseudostreamers. There are also patches of dense and slow solar wind extending away from the HCS.\\

One can see that the foreground domain is submerged by an intense and extended density enhancement associated with the northern streamer. This is consistent with the foreground subimage (Figure~\ref{fig:ADAPT_3imgs}(b)) as \textit{PSP} is close in space to this high-density region. The instrument records a significant increase in brightness over a broad region extending northwards from near the equator. \textit{PSP} is therefore imaging different regions of the streamer from a vantage point that is just below the HPS. \\ 

We confirm that the pseudostreamer rays (green arrow in Figure \ref{fig:WLSynthISPR}) originate near the unipolar cusp (Figure~\ref{fig:PFSSextrap}). This region is located south of the HCS and well in front of the Thomson sphere. Figure~\ref{fig:CarrModels}(a) reveals that the wind forming in that region is not as dense as the simulated streamer flows and would indeed appear less bright in the images. Consequently, this less dense region appears in the foreground subimage as a much fainter diffuse region in the lower half of the image (see the green arrow in Figure~\ref{fig:ADAPT_3imgs} panel b).\\

On the contrary, the background integration domain covers a region of much lower density with a thin and flat layer of local density enhancement associated with the HPS. Imaging this east-west oriented structure from a larger distance explains why the streamer appears this time as a very thin and bright streamer ray visible in the background subimage (indicated by a red arrow in Figure~\ref{fig:ADAPT_3imgs}(c)). \\
\clearpage

Therefore, the full synthetic image (Figure~\ref{fig:ADAPT_3imgs}(a)) shows both the broad and diffuse light scattering emission of the foreground as well as the thin ray of the same northern streamer that flattens at lower latitudes behind the Thomson sphere. We must conclude that comparing WISPR-I images by simply taking slices of simulations near the Thomson sphere is inadequate, and a complete analysis of WISPR-I images requires an analysis that integrates foreground and background features. \\ 

\begin{figure}[ht!]
\centering
\includegraphics[scale=0.48]{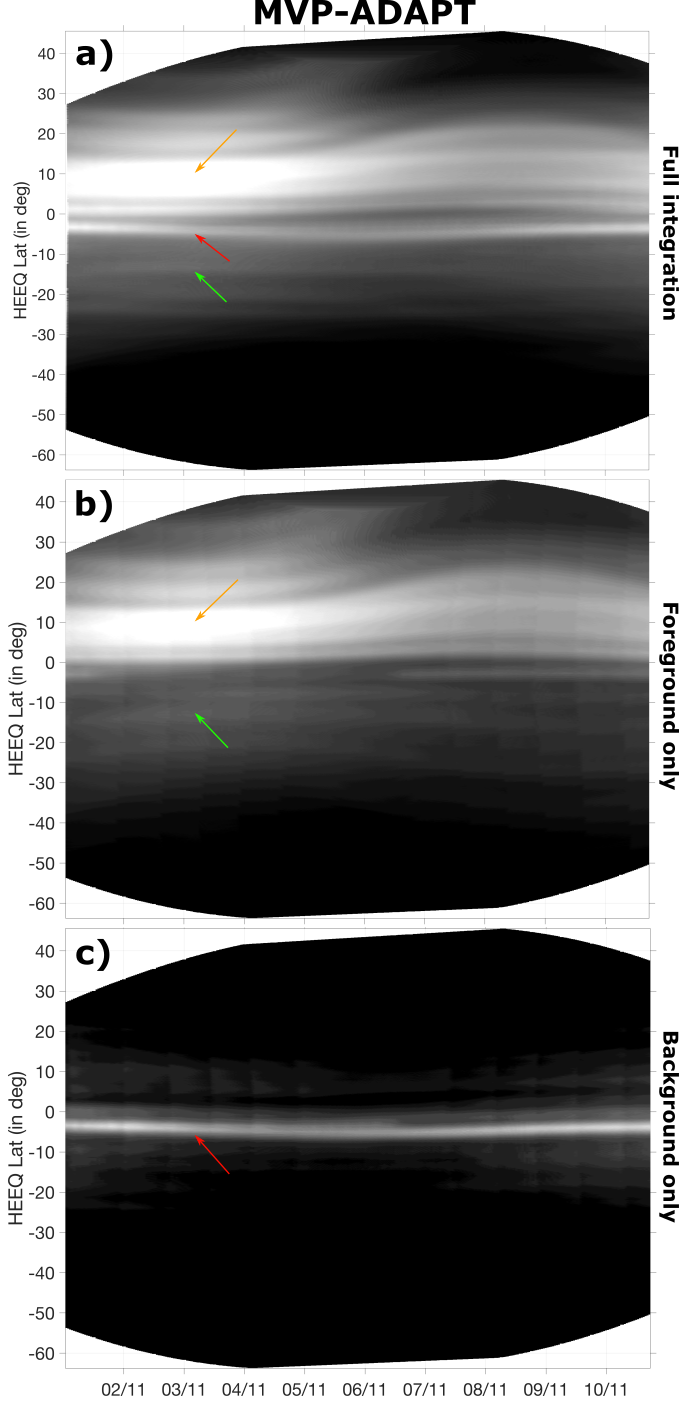}
\caption{Synthetic latitude vs. time maps from the MVP-ADAPT run. They have been generated from (a) the full synthetic images, (b) the foreground subimages only, and (c) the background subimages only. The arrows' color coding is the same as in the previous figures. \label{fig:ADAPT_3maps}}
\end{figure}

Figure~\ref{fig:ADAPT_3maps} shows the MVP-ADAPT map (Figure~\ref{fig:WLSynthCarr}) reprocessed using either the foreground or the background synthetic subimages. Similarly to the analysis done on the WL synthetic subimages, the foreground submap (panel b) contributes to most of the bright structures seen in the full map (panel a). An exception is the thin southern bright stripe already mentioned and located in the background submap (red arrow in panel c). These submaps give us the last hint to understand the origin of the apparent two-lane splitting of the northern streamer ray that we discussed in Section ~\ref{sec:CARR_WISPR}, which is annotated by the orange and red arrows and visible from 2018 November 3 to 9 in the WISPR map (Figure~\ref{fig:WISPR_maps}). \\

From the decomposition of an MVP-ADAPT WL synthetic image into two subimages (foreground and background as shown in Figure~\ref{fig:ADAPT_3imgs}) as well as the analysis of the polarity and density slices of the MVP-ADAPT run, we interpret this splitting as the result of an LOS integration effect from two very distinct regions. The initial northern streamer, slightly folded, visible in the foreground and located at a few degrees above the equator flattens further in the background to remain flat at a few degrees below the equator. This flat part of the streamer in the background is highly visible in the polarity slice (see Figure~\ref{fig:CarrModels}(b)) from $\sim$30$^\circ$ to $\sim$100$^\circ$ Carrington longitude. Therefore, from the modeling, we can identify the apparent streamer splitting in the WISPR map to be the result of an LOS integration effect of a slightly folded HPS extending at and beyond the Thomson sphere. \\

The density and polarity maps (Figure~\ref{fig:CarrModels}(a) and (b)) also reveal that smaller folds of the HCS situated in the foreground could create the rays that drift to higher latitudes away from the brightest streamer rays. This includes the rays located at the highest latitudes (Figure~\ref{fig:ADAPT_3maps}(a)). Several dense flux tubes can also be observed in the southern region in the density map between Carrington longitudes 320$^\circ$ and 340$^\circ$. They are isolated flux tubes with higher densities that produce additional thin rays in the southern part of the synthetic foreground image (Figure \ref{fig:ADAPT_3imgs}). This produces a structuring of the modeled southern rays in the region of the pseudostreamer (Figure~\ref{fig:ADAPT_3maps}(a)) that is also seen in individual WISPR-I images (Figure \ref{fig:SeqWISPR}). These southern rays also drift to higher southern latitudes during the period of super-rotation as seen in the WISPR-I map (Figure \ref{fig:WISPR_maps}). This is due to the proximity of these rays to WISPR-I combined with the motion of the spacecraft plunging toward these rays.\\

In this section, we have exploited high-resolution numerical models of the solar wind to explore the origin of the features visible in WISPR-I observations. Comparisons between the WL synthetic images and maps produced from the MVP-WSO and MVP-ADAPT runs show that high-resolution modeling is essential to understand the finest structures present in the WISPR-I observations such as the streamer splitting discussed just before. The MVP-WSO simulation provides density distributions for the large-scale structures such as the streamer and the pseudostreamer. \\

There is a high consistency between the MVP-WSO and WISPR-I images. MVP-ADAPT unveils very fine features, such as a splitting of the northern streamer into two separate streamer rays as well as the effect of HPS folding, the latter being directly visible in the WISPR observations but not the former. Both MVP-WSO and MVP-ADAPT maps reveal features consistent with the WISPR map. However, both simulations produce features that remain visible over the whole time interval of the encounter that are not as persistent as in the WISPR map. This is a hint that the simulations do not reproduce exactly the extent and location of the HPS and streamers. As a consequence, images from the WISPR-I telescope using a similar analysis carried out in this paper can be a very good test to better constrain and improve numerical models of the solar wind and corona.

\section{Discussion} \label{sec:Discussion}

Modeling of the solar wind and corona has been extensively used in this study not to perform direct comparisons with WISPR images but to help us understand the origin of the different coronal rays observed by WISPR-I during the first encounter of \textit{PSP}.\\

Despite an angular resolution greater than 2$^\circ$ in latitude, both MVP-WSO and MVP-ADAPT simulations appear fine enough to reproduce at least the large-scale features observed by WISPR-I. The reason is that the simulated brightness along each LOS is the result of integrating density over many distinct cells of the MHD cube. Figure~\ref{fig:nlos} illustrates this point with synthetic images from the MVP-ADAPT simulation that have been constructed with a different number of integrating points along the LOS. \\

We showed the need of having a fine-enough simulation (e.g. MVP-ADAPT) in order to reproduce even the smallest features observed by WISPR-I. We could not have identified in our synthetic images and maps the apparent two-lane streamer splitting as seen by WISPR-I by using only the lower resolution MVP-WSO run. The WSO resolution of 5$^\circ$ in latitude and longitude is not sufficient to model the smallest folds of the HPS and the associated LOS effects in WISPR-I. \\

The higher-resolution MVP-ADAPT simulation (with 2$^\circ$ angle resolution) allowed us to give further context and potentially explain the apparent splitting of the brightest streamer rays seen by WISPR-I. Our results suggest that this originates from the LOS integration along an extended region where the HPS undergoes a latitude change. Our model shows that the HPS latitude changes from $\sim$10$^\circ$ to $\sim$-5$^\circ$ over a $\sim$60$^\circ$ Carrington longitude span at the region where WISPR observations were made. The effect of such folds in the HPS have been known since LASCO observations to produce separated streamer rays \citep[see][]{Sheeley1997ApJ,Wang1998}. The novelty in WISPR-I observations is to act as a microscope to catch even small latitudinal changes in the HPS, allowing a more detailed evaluation of current coronal models. \\ 

The MVP-ADAPT simulation additionally reveals other finer structuring of the streamer rays, which is not clear in WISPR images. This could be an issue with the modeling and its assumptions, the sensitivity of WISPR-I observations, the way the F-corona is subtracted from the WISPR images, or could be simply a side effect from rescaling the intensity.  Alternatively, \textit{PSP} is perhaps not yet close enough to reveal some of the additional features revealed in the modeling. This will likely be clarified when \textit{PSP} approaches closer to the Sun on its way to 10 \textit{R}$_\odot$. \\

A last striking difference when comparing modeling to observations emerges from the latitude versus time maps (Figure~\ref{fig:WISPR_maps} and~\ref{fig:WLSynthCarr}). The bright features are in general located at similar latitudes compared with observations, but their extent in time is somewhat different. In Sections~\ref{sec:ObsWISPRrays} and~\ref{sec:CARR_WISPR}, we mentioned how the background removal on WISPR-I images can affect the aspect
of the observed streamer and pseudostreamer rays. As a consequence, the observed streamer splitting might also be impacted. The extent in time of the observed WL features shown in Figure~\ref{fig:WISPR_maps} may evolve in future versions of WISPR-I images with improvements of the F-corona background model and calibration procedure. \\

The source surface was set to the standard height of 2.5 \textit{R}$_\odot$. A comparison of the magnetic field measured in situ by \textit{PSP} with the sector structure derived from PFSS suggests that the source-surface height could be slightly lower during that same period \citep{Badman2020ApJ}. The impact of the source-surface radius on the structure of streamers will be the focus of a future study.\\

PFSS extrapolations usually provide a reasonable representation of the global topology of the corona and have been widely used in the past to identify the location of streamers and pseudostreamers \citep[][]{Masson2014ApJ}. But sometimes they differ from other reconstruction techniques such as tomography \citep[][]{Kramar2016FrASS} or full 3D MHD. Despite its limitations, the PFSS model nonetheless provides a consistent interpretation of the nature of the coronal rays observed by WISPR-I. It is not in the scope of this paper to make a comparison of results from distinct coronal magnetic field reconstruction techniques. However, for completeness, we have compared our results with full 3D MHD simulations provided by \citet{Reville2020ApJ} and Predictive Science Inc.\footnote{\url{https://www.predsci.com/portal/home.php}}, which support the present analysis. \\

In this analysis, we used a single static MULTI-VP simulation over the whole encounter spanning about one-third of the solar rotation. As a consequence, the time evolution of the coronal structures over the \textit{PSP} encounter is not captured. This may be an important aspect. The ADAPT magnetograms and consequently our MULTI-VP MHD model can be updated every 6 hr, hence the time variability of the streamers observed by WISPR-I during the first encounter could potentially be captured by the simulations. Time dependence in simulations involves the analysis of a very large amount of data and is left for future studies.\\

\section{Conclusion} \label{sec:Conclusion}

The unprecedented proximity of \textit{PSP} to the Sun allows WISPR to capture in great detail coronal and streamer rays. The region along the WISPR LOS which contributes most to the total brightness becomes narrower as \textit{PSP} approaches the Sun. By exploiting a 3D solar wind model, we showed that most of the contributions to the brightness in WISPR-I images are from plasma released near the HPS that can extend over a broad region situated in front and behind of the Thomson sphere. Nevertheless, images so close to the Sun are less affected by the superposition of numerous bright and diffuse features along the LOS than near 1~au imaging but remain complex to interpret. \\

During the \textit{PSP} first encounter, WISPR-I observed the western part (as viewed from Earth) of the solar corona and recorded a plethora of coronal features. From those first high-resolution WISPR images of coronal rays combined with the detailed modeling performed in this study, we could explain a number of observational features. Our results and conclusions can be summarized as follows:

\begin{itemize}

    \item From a comparison of WISPR to near 1~au LASCO and STA observations during periods where similar regions were observed, we showed that the WISPR instrument acts like a microscope providing a blown-up view of streamers. Compared with coronal observations from 1~au the WISPR-I images unveil finer substructures in the streamer rays.
    
    \item The latitude versus time map constructed with WISPR images was helpful in studying both the global topology of the corona over the complete first encounter and the evolution of the finer-scale structures. For the first time, WISPR unveils fine substructures inside streamer rays that are likely at the origin of the HPS.
    
    \item We have validated the large-scale MULTI-VP simulations that were able to reproduce many features observed by WISPR. The main streamer and pseudostreamer rays have been identified in both simulations and observations. 
     
    \item The thin splitting of the brightest streamer rays in the WISPR map is interpreted as a small folding of the HPS through the LOS. The drifting coronal rays visible on the WISPR map are also observed on the synthetic maps. These features originate from narrow rays situated closer to \textit{PSP}. Their apparent motion can be interpreted as an effect of \textit{PSP}'s super-rotation into these regions at that time. 
    
    \item Finally, we have interpreted the presence of additional rays drifting toward the southern polar regions to localized source regions of dense solar wind forming above the cusp of the pseudostreamer. 
    
\end{itemize}

Patches of dense solar wind form in our simulations over the entire source region of the slow solar wind roughly below 40$^\circ$ of heliographic latitude. This points toward a 'texture' of the solar wind that finds its root in the highly structured nature of the coronal magnetic field. We expect that as \textit{PSP} gets closer to the Sun and the WISPR "microscope" zooms farther, the rays formed by these patches of dense solar wind will become more apparent from the rest of the corona. The presence of such fine rays have been shown in highly processed eclipse images \citep[see][]{November1996ApJ,Druckmuller2014ApJ}. \textit{PSP} finally provides a way to study these structures systematically and to determine the global structure of the slow solar wind.

\acknowledgments

The IRAP team acknowledges support from the French space agency (Centre National des Etudes Spatiales; CNES; \url{https://cnes.fr/fr}) that funds the plasma physics data center (Centre de Données de la Physique des Plasmas; CDPP; \url{http://cdpp.eu/}), the Multi Experiment Data \& Operation Center (MEDOC; \url{https://idoc.ias.u-psud.fr/MEDOC}), and the space weather team in Toulouse (Solar-Terrestrial Observations and Modelling Service; STORMS; \url{https://stormsweb.irap.omp.eu/}). This includes funding for the data mining tools AMDA (\url{http://amda.cdpp.eu/}), CLWEB (\url{clweb.cesr.fr/}), and the propagation tool (\url{http://propagationtool.cdpp.eu}). The numerical simulations of this study were performed using HPC resources from CALMIP (grant P1504). A.K. acknowledges financial support from the ANR project SLOW{\_}\,SOURCE (ANR-18-ERC1-0006-01), COROSHOCK (ANR-17-CE31-0006-01), and the FP7 HELCATS project \url{https://www.helcats-fp7.eu/} under the FP7 EU contract number 606692. The work of A.P.R., N.P., L.G., V.R. was funded by the ERC SLOW{\_}\,SOURCE project (SLOW{\_}\,SOURCE - DLV-819189).  The work of P.H., R.A.H., N.R., G.S., A.T. and A.V. was supported by the \textit{PSP}/WISPR program. We thank the \textit{PSP}-WISPR, \textit{STEREO}-SECCHI and \textit{SOHO}-LASCO teams.

\appendix

\begin{figure}[h!]
\centering
\includegraphics[scale=0.68]{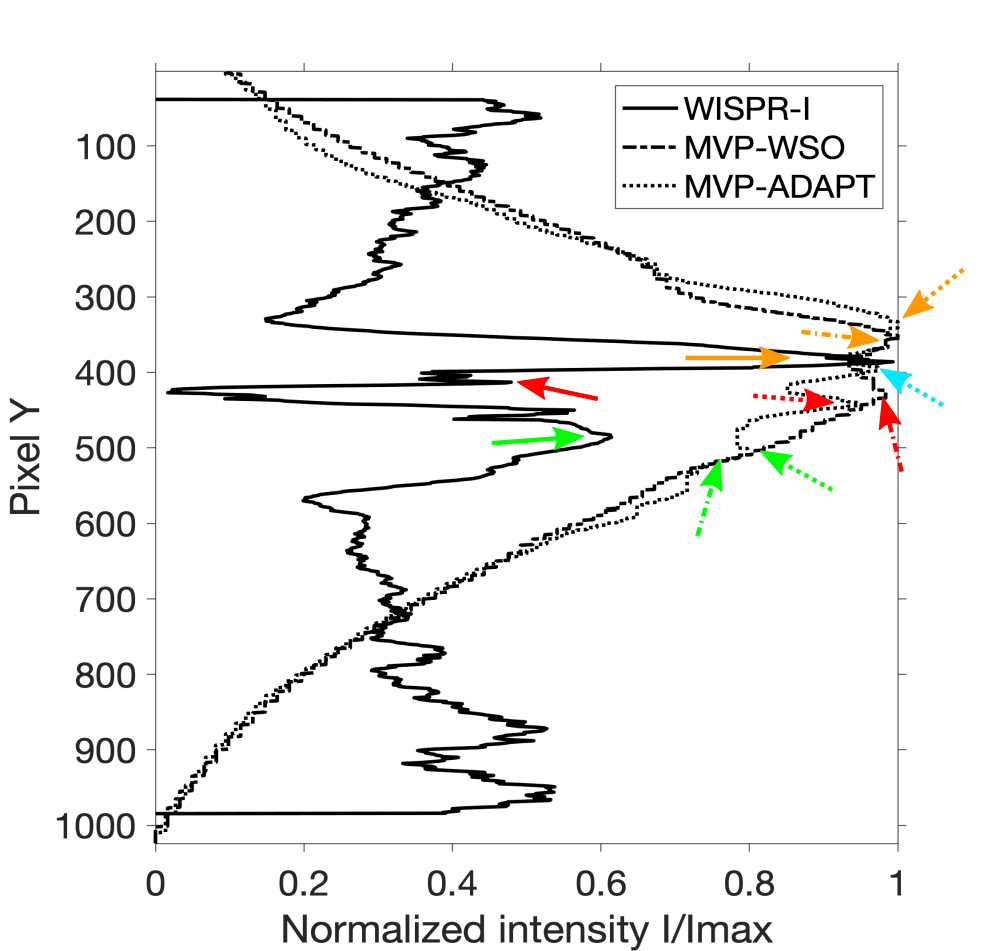}
\caption{Normalized intensity distribution along the vertical slices depicted as green lines in Figure~\ref{fig:WLSynthISPR}. The arrows color coding is the same as in Figure~\ref{fig:WLSynthISPR}. \label{fig:vert_slice}}
\end{figure}

\newpage

\begin{figure}[h!]
\centering
\includegraphics[scale=0.05]{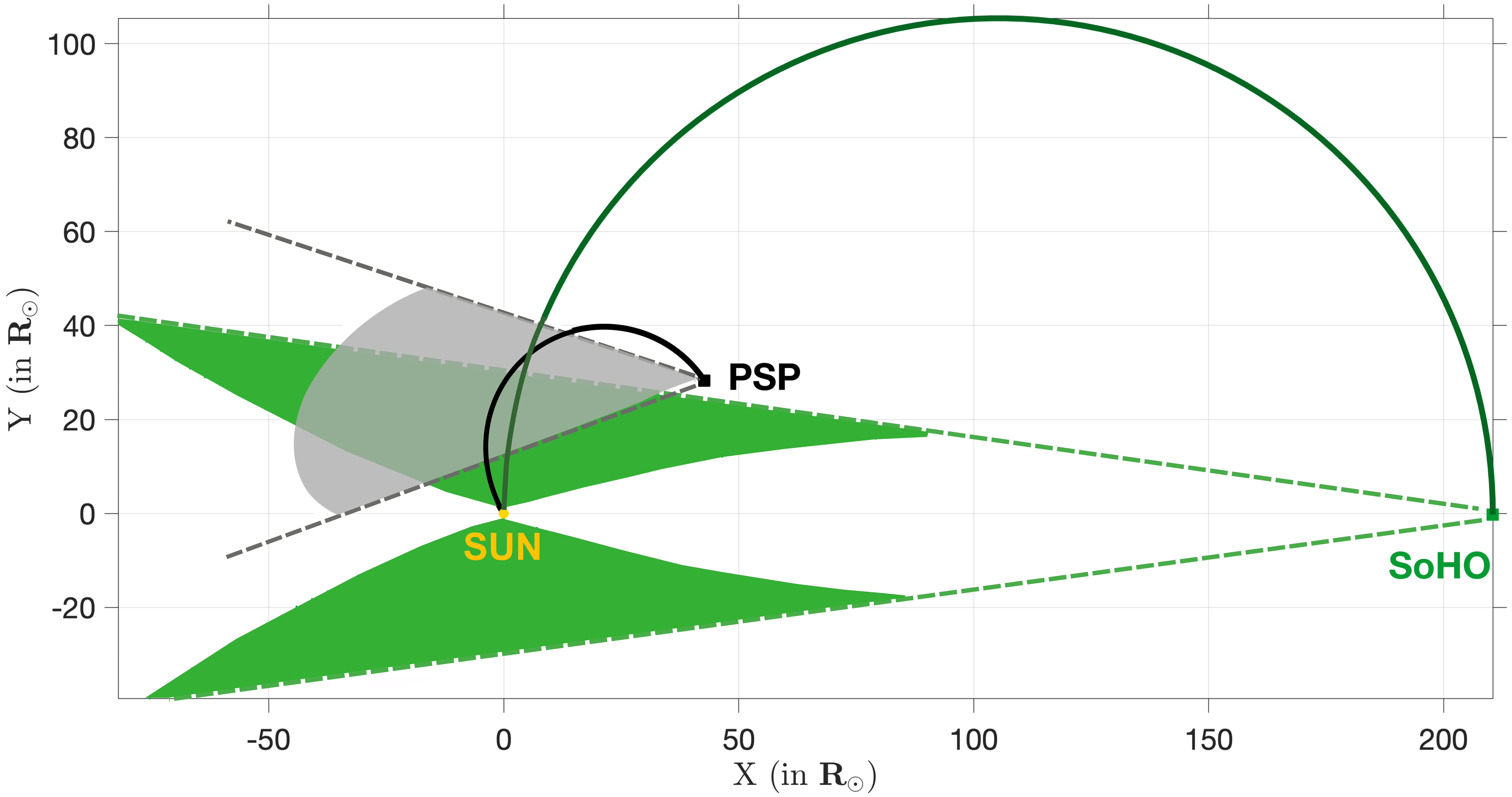}
\caption{View of the ecliptic from solar north showing the position and FOV of \textit{PSP} and \textit{SOHO} on 2018-11-01 at $\sim$00:00~UT. The Thomson spheres for \textit{PSP} and \textit{SOHO} are represented in black and green respectively. The region contributing to 99\% of the total brightness received by WISPR-I and LASCO-C3 are shown in grey and green respectively. \label{fig:Thomson_fig}}
\end{figure}

\begin{figure}[h!]
\centering
\includegraphics[scale=0.5]{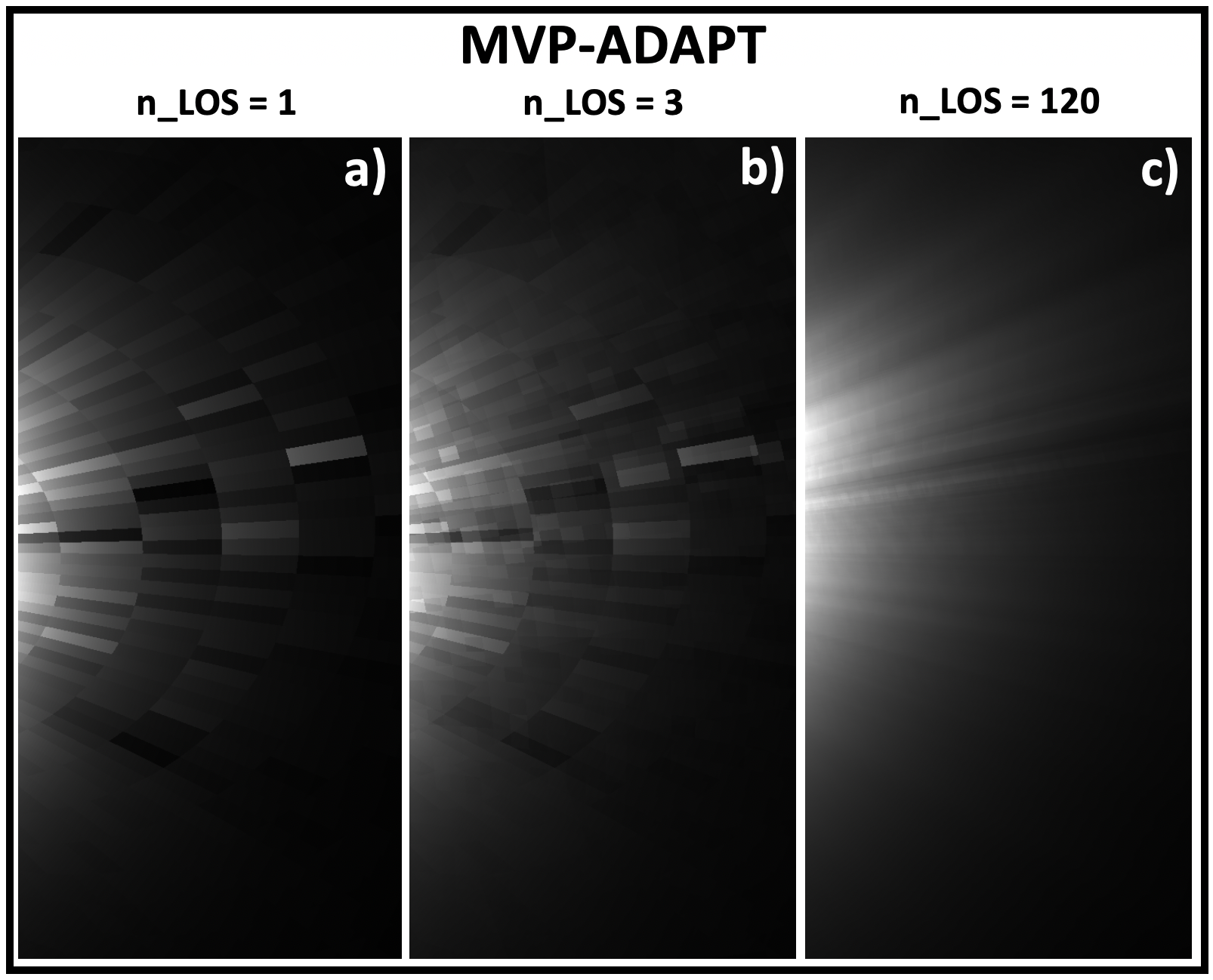}
\caption{Synthetic images from the MVP-ADAPT run. In each image we consider a different number of cells to the construction of the integrating LOS. Panels a, b and c correspond to LOS that are made of 1, 3 and 120 integrating points respectively. \label{fig:nlos}}
\end{figure}

\newpage

\section{Supplemental movie}
\label{sec:Appendix}

\begin{figure}[h!]
\centering
\includegraphics[scale=0.85]{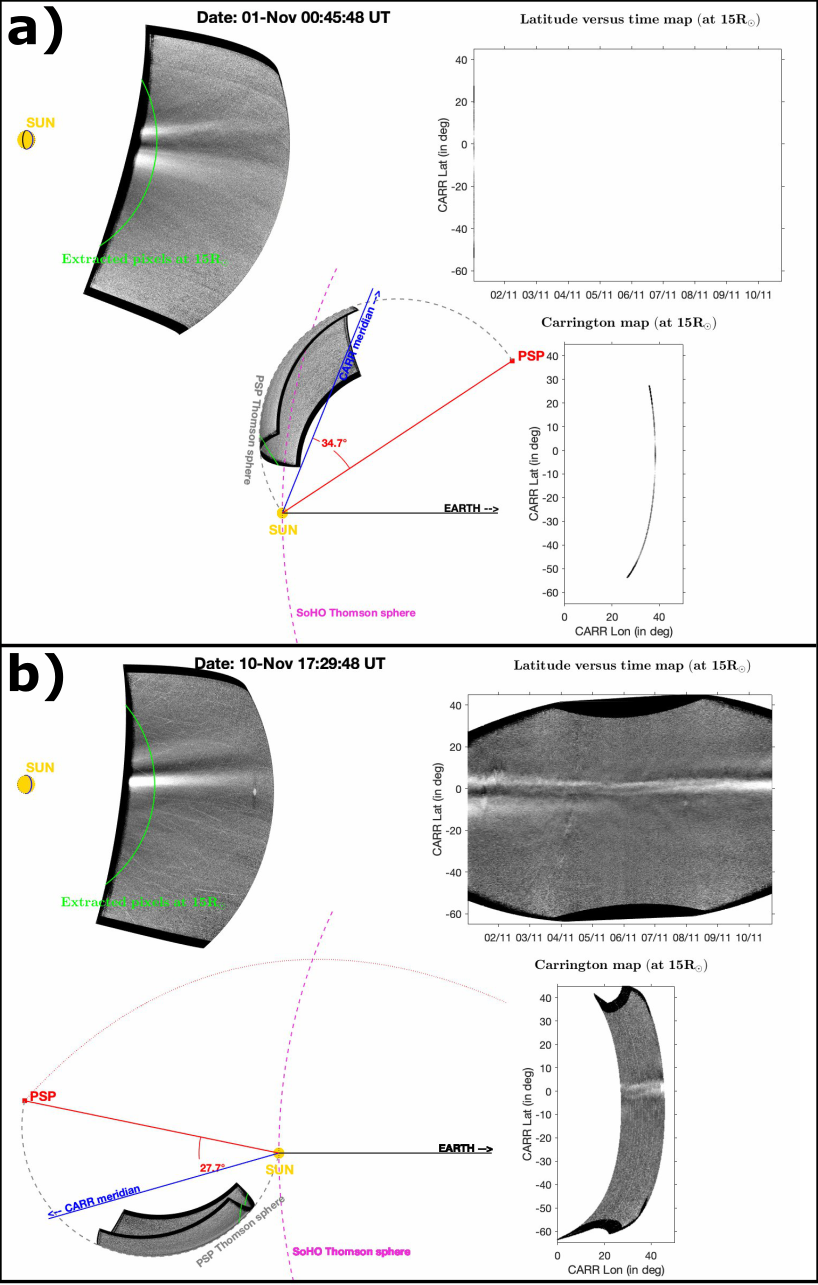}
\caption{First (a) and last (b) frames from the \href{https://nuage.irap.omp.eu/index.php/s/x2tHzaBM9fyL1bt}{supplementary movie}.
The upper-left panel is a 3D view from the \textit{PSP} position of the pixels that are extracted at 15 R$_{\odot}$ from the WISPR-I images. These extracted pixels are mapped in a Carrington latitude vs. time format (upper-right panel) and in a Carrington latitude vs. longitude format (lower-right panel). The lower-left panel shows from solar north the orbital configuration of \textit{PSP} during the first encounter. \label{fig:frames_movie}}
\end{figure}

A \href{https://nuage.irap.omp.eu/index.php/s/x2tHzaBM9fyL1bt}{supplementary movie} shows how the WISPR-I Carrington map is built over time (Figure~\ref{fig:frames_movie}). The video begins on 2018 November 1 at 00:45~UT and ends on 2018 November 10 at 17:29~UT.

\end{document}